\documentclass[1p,authoryear,times]{elsarticle}

\usepackage{graphicx}

\usepackage{amssymb}
\usepackage{lineno}
\usepackage{lscape}
\usepackage{tabularx}
\usepackage[gen]{eurosym}
\usepackage{multirow}
\usepackage{hhline}
\usepackage{booktabs}
\usepackage{siunitx}
\newcolumntype{L}{@{}l@{}} 
\newcommand{\mc}[1]{\multicolumn{1}{c}{#1}} 

\journal{The Journal of Systems and Software}

\begin{document}

\begin{frontmatter}

\title{Group development and group maturity \\when building agile teams:\\ A qualitative and quantitative investigation at eight large companies}



\author[torkar-new]{Lucas Gren\corref{cor1}}
\ead{lucas.gren@cse.gu.se}
\cortext[cor1]{Corresponding author. Tel.: +46 739 882 010}

\author[torkar-new,torkar]{Richard Torkar}
\ead{richard.torkar@cse.gu.se}

\author[torkar,torkar-new]{Robert Feldt}
\ead{robert.feldt@bth.se}

\address[torkar-new]{Chalmers University of Technology and the University of Gothenburg, SE-412 96 Gothenburg, Sweden}
\address[torkar]{Blekinge Institute of Technology, SE-371 79 Karlskrona, Sweden}

\begin{abstract}
The agile approach to projects focuses more on close-knit teams than traditional waterfall projects, which means that aspects of group maturity become even more important. This psychological aspect is not much researched in connection to the building of an ``agile team.'' The purpose of this study is to investigate how building agile teams is connected to a group development model taken from social psychology. We conducted ten semi-structured interviews with coaches, Scrum Masters, and managers responsible for the agile process from seven different companies, and collected survey data from 66 group-members from four companies (a total of eight different companies). The survey included an agile measurement tool and the one part of the Group Development Questionnaire. The results show that the practitioners define group developmental aspects as key factors to a successful agile transition. Also, the quantitative measurement of agility was significantly correlated to the group maturity measurement. We conclude that adding these psychological aspects to the description of the ``agile team'' could increase the understanding of agility and partly help define an ``agile team.'' We propose that future work should develop specific guidelines for how software development teams at different maturity levels might adopt agile principles and practices differently. 
\end{abstract}


\end{frontmatter}

\section{Introduction}\label{intro}
Groups have existed as long as humans and our ability to form and work in groups is key to our survival and development. However, some people dislike working in groups because group work can be cumbersome and involve conflict, hurt feelings, and inefficiency. The reason why organizations want to organize work in group-form is because when a group is working well, it works extremely well compared to other work methods~\citep{wheelan2012}. This aspect has been evident in software engineering since the beginning of the field (see e.g.~\citet{weinberg}), but more focus on these aspect have lately been called for by some researchers~\citep{lenberg2015}. One factor of why software projects failed that is often mentioned has been the traditional approach to software development where projects, usually, were considered to be ``plan-driven''~\citep{petersen2}. These methods come from systems engineering and other disciplines, and were established to coordinate large inter-operating components. However, software does not function as hardware and different standards were therefore introduced~\citep{boehmm}. These new standards include many aspects of psychology since they are based on human interaction in order to deliver customer value faster~\citep{adkins}.


When changing to an agile method (e.g.\ eXtreme Programming, Scrum, Lean etc.), where cooperation and a self-organizing team are central, some aspects of the modern workplace might cause problems. If group members are unable to, e.g.\ be physically present during meetings, the aspect of human interaction becomes harder to achieve and problems concerning communication, culture, trust, and knowledge management appear~\citep{jala}. Because of the agile management technique, organizational psychology issues have gotten more attention in software engineering~\citep{bali,lenberg2015}. This research aims at explaining group psychological (and especially group developmental) aspects of building agile teams that could help in understanding why some agile transitions succeed and some do not. 

Over the years there have been many models within psychology on how groups behave. There seems to be a common pattern of what happens to all human groups regardless of different sectors or where they are located in the world \citep{wheelan2005}. However, this is one of the first studies that investigate group development in the software engineering domain, but since the theory has been shown to be valid in most other fields we see no reason why such a core concept of human behavior would not be present in the software engineering domain. The patterns have been categorized into different stages and labeled differently by many researchers. \citet{bion}, for example, states that a group always has two states; the work group, and the basic assumption group (consisting of dependency, fight-flight, and pairing stages). \citet{tuckman} defined a classic development model with the phases; forming, storming, norming, and performing. These stages perfectly correspond to the theory used in this study and are described in the Section~\ref{sec:related_work} in this paper. The reason behind this choice, was that it is an integrated model of group development built on an extensive body of research. In addition, it is also the only evidence-based group development model known to date \citep{wheelan}. 

The study of psychological aspects of agile development is quite a new research field and some studies have been conducted regarding agile methods in connection to culture~\citep{iivari, tolfo2008,whit,tolfo}, personality traits \citep{mcdonald,seger,feldt2010}, and job satisfaction~\citep{melnik2,gren1}, but only one article has been found on agile work-groups and group psychology~\citep{teh}, in which they conclude that productive group norms give better results. One issue that often surfaces in the modern software development workplace is collocation. All teams need to mitigate challenges connected to being geographically spread out in the same way. The solutions suggested by \citet{noll2010global} are ``site visits, synchronous communication technology, and knowledge sharing infrastructure to capture implicit knowledge and make it explicit.''

The relationships between people in groups and aspects of group maturity have been shown to have large effects on effectiveness, in fact, mature groups have been shown to perform much better in a diversity of fields, e.g.\ they finish projects faster~\citep{wheelan1998}. Students perform better on standardized tests if the faculty work-group is at a mature development stage~\citep{wheelan1999, wheelan2005}, and intensive care staff functioning in a mature work-group save more lives~\citep{wheelan20032}. All these studies have shown that paying attention to group development could help the group to increase its performance and therefore, in the end, provide a higher rate of what is considered project success. In the agile development domain, an agile approach to projects has been shown to be more successful~\citep{serrador2015does}, which gives rise to the question of if group maturity could be one key aspect of this difference in success. 

In order to clarify the concepts studied in this paper, we use the term ``group development'' to refer to the developmental process of getting a work-group to mature over time (also sometimes called the group maturity level). The difference is made clear in Section~\ref{sub:integratedgroup}, but groups that score higher on the measurements of the later development stages are considered more mature than groups that score lower on these scales. ``Performance'' is connected to group maturity since more mature teams have been shown to perform better as compared to other teams, as described earlier. The concept of ``success'' is twofold according to \citet{de1988measurement}, one part being ``project success,'' which refers to an evaluation against project criteria, and the other being ``project management success,'' which includes performance.  Since agility has been shown to increase project success in software engineering and group maturity has been shown to be connected to group performance, we believe that looking at the connections between group maturity and agile teams, in the long run, could provide helpful guidance from group development psychology to agile project management. 

The main contribution of this study is an in-depth analysis of qualitative data showing how a lot of the agile coaches, Scrum Masters, and managers work with group development issues even though the developmental perspective of groups is not explicitly described and empirically tested in the agile literature. Furthermore, we collected survey data from 66 employees and correlated a measurement of agility to that of group development and found supporting evidence of the importance of adding the group development perspective to the agile team models.

\subsection{Research Question}
This study has the following research question:
\begin{itemize}
\item ``How is group maturity connected to building agile teams?''
\end{itemize}

To answer this question we did a diversity of analyzes, both with interviews and surveys to investigate the connections and discuss how the agile approach would benefit from adding the group developmental dimension to its implementation theories. 

\paragraph{Earlier Research and Publications}
The issue of trying to measure group development and correlate it to agility has been a continuous work for the authors of this paper. The first hurdle was to find a tool that could be considered as a valid measurement of agility; but how can one measure something that is basically undefined and means different things to both researchers and practitioners? We have not found many thoroughly validated agile measurements, and we selected one of the well-cited tools to measure ``agility,'' based on the overview presented by \citet{lepp}, and a validation study conducted by \citet{ozcan}. As mentioned, one issue is the definition of agility since we need to know what to measure. The reason why we chose Sidky's~(\citeyear{sidkyphd}) tool is that it provides a set of items in survey form that aims to measure the behavior connected to agile processes instead of having participants tick what practices they use from a list. We have also published a separate validation study on that tool \citep{grenjss}, but our conclusions from that study was that much work is needed to claim that such a tool measures aspects of ``agility.'' Also, an overall correlation analysis between the group development questionnaire, developed by \citet{wheelan}, and agility, as defined by \citet{sidkyphd}, was published in \citet{gren2015seaa}. All in all, the previous work showed some connections between the concepts but gave us little insight of how\slash if the agile practitioners work on group development when trying to build agile teams. 

Therefore, this study employes a different research strategy. Since the measurement of agility has been shown to be contextually dependent and tricky to measure quantitatively, we tried to find out how practitioners work on building agile teams in connection to group development through in-depth qualitative data gathered through interviews. In order to triangulate the concept we still collected additional survey data and ran a correlation analysis of the agile categories as developed in \citet{grenjss} and the group development questionnaire Scale 4 mean values (see our work in \citet{gren2015seaa}).

Since we do not know details about the connections between group development and agility this study is exploratory in its nature and we aim at describing the connections we find in both the qualitative and the quantitative data.

Section~\ref{sec:related_work} will outline group development research and present the agility measurement used, Section~\ref{sec:methodology} will present the methodology used in this paper, Section~\ref{sec:results} will present the interview summaries and interpretation of the qualitative data, but also survey results and the statistical tests conducted. Section~\ref{sec:discussion} will discuss results, and, finally, Section~\ref{sec:conclusions} will present conclusions and suggest future work.

\section{Related Work}\label{sec:related_work}

\subsection{Groups and Teams}
\citet{grupp} defines a group as: ``three or more members that interact with each other to perform a number of tasks and achieve a set of common goals,'' which means that large groups are in fact a set of smaller subgroups and should be considered separately. If a group consists of more than eight individuals it is less productive than a smaller group~\citep{wheelan2009}. A ``work group'' consists of members that want to create a shared view of group goals and develop a structure to achieve these goals. A ``team'' is a work group that has shared goals and effective methods to achieve them~\citep{wheelan}. This implicates that many work groups in organizations are not teams, and only 17\% of all groups were considered teams in a study by~\citet{wheelan}. We have not found any more recent studies that surveyed a large number of teams, but possibly, the team-based work focus in the last decades might have increased this percentage, but we have not found any empirical support for such a statement. Nor have we found any previous studies on group development in software engineering, which means that we do not have an estimate of how mature software engineering teams are in relation to other fields.

\subsection{Wheelan's Integrated Model of Group Development}\label{sub:integratedgroup}
\begin{figure*}
\centerline{\includegraphics[width=130mm]{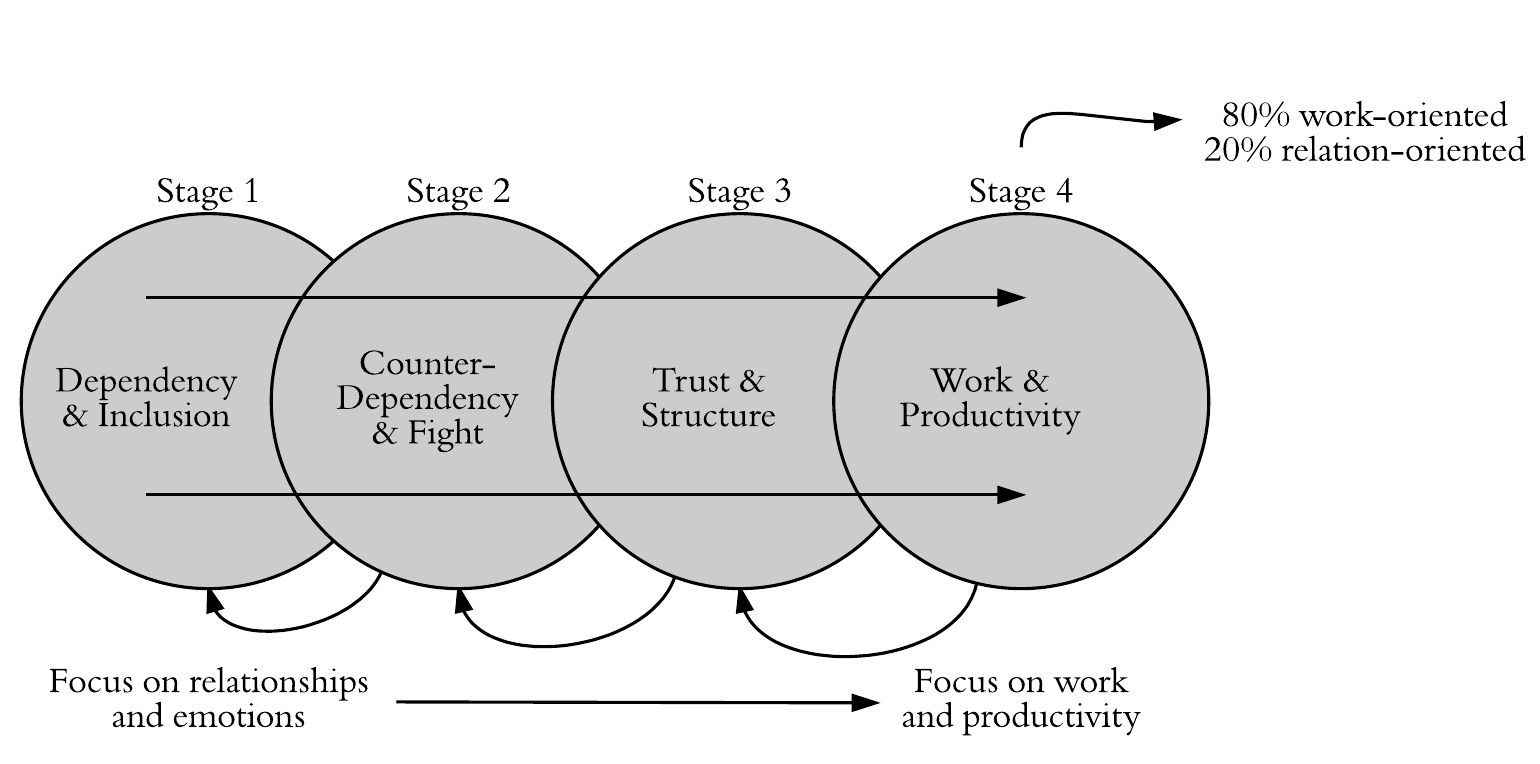}}
\caption{The Group Development Stages (adopted from \citet{wheelan2012})}
\label{fig:groupstages}
\end{figure*}

Many group development theories describe a dynamic view of the group. Older theories, like the one presented by \citet{bion}, as well as newer group dynamic theory all evolve around a set of stages that groups go through over time~\citep{wheelan1993}. The theory used in this study presented by~\citet{wheelan} is actually an integrated model of group development and is also branded as such (the model is called The Integrated Model of Group Development, or IMGD). This model claims to be cyclic but integrates all other four different types of models (sequential, life-cycle, equilibrium, and adaptive models) and sees these theories as stemming from differences in group types, group tasks, time the groups had met together, lack of clarity of group stages and phases, and issues of group progress versus reoccurring themes \citep{wheelandev}.
\citet{wheelan} later connected a survey to this model, called the Group Development Questionnaire (or GDQ). This tool measures the maturity level of a group in four different stages (see Figure~\ref{fig:groupstages}). These four stages which will be presented in more detail next and the Group Development Questionnaire will be explained in detail afterward.

The first stage (Stage 1: Dependency and Inclusion or ``Forming'') consists of three main areas: ``concerns about safety and inclusion,'' ``member dependency on the designated leader,'' and ``a wish for order and structure.'' There is a set of things the group have to do to fulfill these purposes, and the first part is to create a sense of belonging and create a foundation of predictable patterns of interaction. The behaviors and feelings in the first stage are categorized by concern about personal safety in the group, members seek to be accepted by other members and the leader, they fear rejection, and they communicate in tentative and polite ways. Another aspect is the dependency on the designated leader. Members will express a need for dependable and directive leadership in the first stage. The members will view the leader as benevolent and competent and expect the leader to provide direction and personal safety. At this stage the leader is rarely challenged and is accepted as the leader by the group by the group members. Cohesion and commitment to the group will be based on identification with the leader. In the beginning goals are not clear to members, but clarification is not sought. Role assignment is often based on external status and first impressions rather than real competence with goal and task requirements. Most of the communication goes through the leader and participation in the discussion is often limited to a few extroverted members. There is also an evident lack of organization and group structure but conflict is minimal and people seem to agree on most topics. Deviation from emerging norms is rare and so are subgroups and coalitions \citep{wheelan}.

The second stage (Stage 2: Counter-Dependency and Fight or ``Storming'') is a stage with conflict where such disagreements must occur in order to create clearer roles, and for the group to create a structure needed to be able to be constructive in the way the group members work together. In this second phase, hard work is needed to get though the conflicts, because shared views of values, norms, and goals need to be put in place. In stage 2, extensive work is needed to agree on these aspects and every member needs to participate for this to happen. The group must get through conflict (opposition between ideas etc.) to develop a unified set of goals, values, and procedures. These conflicts happen because differences of opinion regarding the rules of the game (i.e.\ group norms) are very likely to occur. Conflict is also necessary to build trust. Only by putting efficient conflict resolution in place and by working on finding a unified culture, can the group collaborate well. When a group begins the second stage it is likely that they show dissatisfaction with roles and start to clarify them. Group members also show disagreement about strategies to achieve task accomplishment and about the decision-making process. In the end of the second stage there will be an increased consensus about goals and culture, and conflict resolution, if successful, will increase trust and cohesion in the group \citep{wheelan2012}.

The third stage (Stage 3: Trust and Structure or ``Norming'') develops a structure for the group where the roles are based on competence instead of the initial need for safety or power. The communication is more open and focused on the tasks. The third phase is characterized by further mature negotiations about processes, roles, and the group's organization. When the groups are negotiating roles, organization, and processes, the goals will be much clearer and the group members will agree on them to a larger extent. Also the roles and tasks are adjusted to increase the probability of goal achievement. The role of the leader is more consultative and less directive in this phase and the communication structure is more flexible. Task-oriented instead of relation-oriented content in the communication is also visible here, and a greater tolerance for subgroups, cliques, and coalition is shown. Labor is also divided better between the group members. The conflicts do not disappear but instead continue to occur, but the difference is that they are managed more effectively. When the positive relationships are built in the group there is an increase in trust and group cohesion. Individual commitment to group goals and task is high and voluntary conformity with the group norms upsurges. Deviation from the group will be accepted if considered helpful to the group \citep{wheelan2012}.

The fourth and final stage (Stage 4: Work and Productivity or ``Performing'') is when the group focuses on getting the job done well at the same time as the group cohesion is maintained over a long period of time. Teams that work 80\% and put 20\% into dealing with conflict and intra-personal issues are the most effective ones. The team also encourages task related conflicts and focuses on decision making. If a group reaches stage 4, and therefore becomes a team, they become immensely productive as well as effective. Work is done at all stages in group development, but in stage 4 there is a large increase in focus on task accomplishment. Excitement in enjoying work is often present and an ease in getting work done. Members of a team are usually thrilled and want stay working in the same way for as long time as possible. The norms usually include encouragement of high performance and quality and the team expects to be successful. The team will also encourage innovation and pay attention to details of its work. Making decisions is a careful process at stage four that involves time spent on defining problems or decisions the group must solve or take. The team also spends time on planning how to do this and discussing the actual problems and decision before acting. The team will define decision-making methods that are participatory, and implement and evaluate its solutions and decisions. Periods of conflict are frequent but brief because the team has developed effective conflict management strategies. By getting, giving, and utilizing feedback about its effectiveness and productivity, the team can maintain its high performance. This is also done by evaluating the performance regularly. The team also takes measures to avoid getting stuck in a routine \citep{wheelan2012}. 

According to \citet{sundstrom1990} and \citet{guzzo1985}, the most effective interventions are goal setting and feedback that includes attention to group development issues adapted to the current group stage they are in. Groups can also pend between different group development stages and when members change the group needs to rework parts of the development. This is also dependent on how many new members join and how many quit. One new member is easier to integrate than half of the members \citep{wheelandev}. 

Some research has been conducted in software engineering regarding developers and personality traits (see e.g.\ \citet{mcdonald,seger,feldt2010}). However, a recent mapping study analyzing 40 years of personality research in software engineering, shows no congruent results \citep{forty}. This could be due to the fact that personality traits have been shown to vary over time \citep{terracciano2005hierarchical} and people can change their personality traits \citep{hudson2015volitional}. Since people act differently depending on the level of their group's maturity, we believe the group developmental perspective might add to the understanding of the behavior shown by members of agile software development teams.

\subsection{The Group Development Questionnaire (GDQ)}
In our opinion, the largest contribution by~\citet{wheelan2012} was to connect a questionnaire to group development. In doing so it has become possible to diagnose and pinpoint the current maturity level of a group. Their survey has a total of 60 items and provides a powerful tool for research and interventions in teams. The Group Development Questionnaire is divided into four different parts. Each of them measure how much energy is put into each stage of group development. The fourth part (GDQ4) measures work and productivity and has been shown to correlate with a diversity of effectiveness measures in different sectors, e.g.\ groups that have high scores on GDQ4 finish projects faster~\citep{wheelan1998}, students perform better on standardized test if the faculty team has high scores on GDQ4~\citep{wheelan1999, wheelan2005}, and intensive care staff saves more lives~\citep{wheelan20032}, i.e.\ the measurement seems to be valid across fields.

\subsection{Agility Measurement}

When looking for a way to measure agility we believe a value-driven tool is better than methods suggested by \citet{kurapati} or \citet{korhonen}, because their methods let individuals tick what practices they use, instead of asking about behavior connected to the agile practices that implements the agile principles. This means that even individuals without knowledge of agile terms can reply to the survey. However, as mentioned in Section~\ref{intro}, a fundamental problem of agile maturity measurement models is that they are not scientifically validated \citep{lepp}. 

Some research show that a software engineering process change is far from clear to the organization. Also, the cultural changes described in the Agile Manifesto and research conducted by e.g.\ \citet{doingtobeing,tolfo} shows that agility has both concrete practices, but that these can be implemented without the desired effects, if the culture is not also changed on a deeper level. Cultural changes in organizations imply behavioral changes that take time and the aspect of face-to-face communication stated as utterly important in the study by \citet{williams}, shows that agility is very hard to transition to if people do not meet face-to-face. This also shows the complexity of introducing such methods and habits, core values, beliefs, priorities, politics, attitudes, perceptions, and assumptions often take a huge effort to change \citep{kotter2007leading}.

\citet{ozcan} somewhat validated different agile maturity models on six different aspects, namely fitness for purpose, completeness, definition of agile levels, objectivity, correctness, and consistency, and Sidky's (\citeyear{sidkyphd}) Agile Adoption Framework was given the best result in their study. \citet{jalali2014} also showed that the same agile measurements give different results with practitioners, a result that was also shown by \citet{kostasxp} where they tested three different tools for convergent validity. These different tools, stated to measure the same agile practices, gave totally different results with exactly the same teams. This motives our qualitative approach to assessing agility and its connection to group development. 

We would like, again, to highlight the issue with measuring agility since it is an ambiguous construct, and in a very recent study by \citet{dikert2016challenges} they show that scientific studies on large-scale agile transformation are rare. In addition, quantitative measurements of agile transformations tend to only focus on external hard metrics and do not explore the key human factors of successful transitions (see for example \citet{olszewska2016quantitatively}). Agile maturity or agile maturity levels are, of course, then also very difficult to assess. However, we believe that some behavior could be considered more ``agile'' than the more traditional approaches to projects. This spectrum of agile behavior is how we define agile maturity or ``levels'' of agility in this paper.

Sidky's (\citeyear{sidkyphd}) Agile Adoption Framework defines which agile methods an organization is ready to use based in its agile potential. The other existing methods are not usable for generalizations according to Sidky since they are merely anecdotal success stories. \citet{boehm20} presented a framework that is criticized by Sidky since they assess agile methods in its generic form and not the practices used. Sidky's framework is divided into ``agile levels,'' ``principles,'' ``practices and concepts,'' and ``items or indicators.'' The agile level is a set of practices that are related to each other and leads to the realization of a core agile value. The agile principles are taken from the agile manifesto and are needed to make sure that the development process is agile (the principles used are derived from the basic and common concepts of all agile methods, since they are taken from the definition of agility). The agile practices and concepts are methods that can be used in the agile processes.

Recently, \citet{grenjss} partly validated the Agile Adoption Framework and suggest new categories for Sidky's (\citeyear{sidkyphd}) items. These new categories were used in the data analysis in this paper (see \citet{grenjss} for more details). The items used in this study is also presented in the Section~\ref{sec:methodology} of this paper. However, because of the mentioned difficulty in measuring agility quantitatively we only use this result as an addition to our in-depth interview analysis of the agility and group development relationship.

\section{Method}\label{sec:methodology}
This section presents the method used to assess agility, group maturity, and correlate these measurements.

\subsection{Participants}
This study was conducted with SAP America Inc.\ and they mediated most of the contacts. The contacts were mediated through an internal experience forum online but the researchers did not select certain teams nor had any relation to the participants. A detailed table of the participant and what type of data collected from each company is shown in Table~\ref{fig:companies}.


\begin{table*}
\renewcommand{\arraystretch}{2}
\caption{Participants in the study.}
\label{fig:companies}
\centering
\begin{tabular}{ccccc||c||c||c||c}
\hline
\bfseries  & Survey Participation (\#) \bfseries & Interview (\#) \\
\hline\hline
Company $A$ & Yes (28 individuals from 3 teams) & Yes (1) \\
\hline
Company $B$ & No & Yes (1)  \\
\hline
Company $C$ & No & Yes (1) \\
\hline
Company $D$ & Yes (17 individuals from 4 teams) & Yes (1)  \\
\hline
Company $E$  & No & Yes (1) \\ 
\hline
Company $F$  & No & Yes (2)  \\
\hline
Company $G$  & Yes (13 individuals from 3 teams) & Yes (3) \\ 
\hline
Company $H$  & Yes (8 individuals from 2 teams) & No \\ 
\hline
\bfseries Total & \bfseries 66 individuals from 12 teams & \bfseries 10 \\
\hline
\end{tabular}
\end{table*}

We will now present some context and company background for the participating companies and interviewees. The company descriptions are taken from the interview transcripts and are the interviewees own descriptions of the context of their work.

\subsubsection{The agile journey at Company $A$}
Company $A$ was using an agile approach within their business and IT projects, such as customer configuration or process improvement within their SAP implementations. Before this approach they had never used a well-defined or strict project management structure, but instead a more schedule-driven or ``loose waterfall'' approach. When planning projects the culture had been that the initial estimates were fixed and committed to. The culture still somewhat had a control mechanism to project management around the agile projects. Their agile journey started with smaller pilot projects, and the intention had been to begin with three agile teams, but they started ten at the same time. At the moment of this research they had ten agile projects running and were using a tailored version of Scrum and Kanban. The biggest challenge at the point of this research was to adhere more to the agile principles and to change more of the organizational thought process, rather than the practices. 

\paragraph{Interviewee at Company $A$} 
The interviewee from Company $A$ was an enterprise agile coach and was leading the entire agile implementation. The interviewee was currently working on changing the mentality of other managers in the organization and getting them not to dictate a date and budget to the projects without involving the team. The interviewee had recently been trying to focus more on the agile principles and the twelve values and reinforcing the Scrum model to the project work across the company. Trying to shift focus from ``what'' to ``why'' they had a certain practice and also to spread the concept of continuous improvement.

\subsubsection{The agile journey at Company $B$}
Company $B$ had its own IT organization that participated in this research. They were working with Product Life-cycle Management (PLM) but had started leveraging an agile approach to their SAP implementations. They had used a strict management approach to these implementations before with a classical Enterprise Resource Planning (ERP) waterfall process with well-defined stages and gates. Their concern was that if they did not get their initial data model right from the beginning the changes became very costly. Therefore, they had started looking at prototyping and the agile approach to projects. They had started with one agile pilot project that they had evaluated afterward. At the time of this research they had four agile projects running in parallel and they had more strongly started to define their agile process with release planning, user stories, and tracking tools. Their Project Management Office (PMO) was currently working on creating an agile methodology with tools that was to be available to the entire organization. Since the agile journey had started with just one low-profile pilot project, they had not been able to involve agile consultants, which was also described as challenging in the beginning. The current challenges were the company ecosystem since they had had to start with one project only, which meant that, to the surrounding organization, the sprint team had been ``just another client.''

\paragraph{Interviewee at Company $B$} 
The interviewee from Company $B$ was a part of the IT organization and was a project manager. The interviewee was currently managing a very large execution project, but was also responsible for the whole ERP delivery at the company. The interviewee was already an experienced project manager but had come across agile during studies both on and off work. The interviewee was the one that had suggested an agile approach to the company and had started the first pilot project, in which a tailored methodology had been tested in order to fit the company ecosystem.

\subsubsection{The agile journey at Company $C$}
The reason for looking at an agile approach at Company $C$ was to deliver value to customers and engage internal and external customers more. Traditionally, they had had more of a waterfall approach to projects but more or less depending on where in the organization. The benefits they saw at the point of the interview were an increasing throughput, shorter cycle time to deliver, and a product that was more in line with what the end-users wanted. They also experienced a better partnership between IT and the business side. At the point of the interview, they had daily stand-ups, release planning, reviews, retrospectives, story grooming and sprint planning. 

\paragraph{Interviewee at Company $C$} 
The interviewee from Company $C$ was the overall responsible for the agile transition working with an executive in order to shepherd that process. They had invested in agile coaches and external experts in order to get started, but then they had conducted and had improved their process internally.

\subsubsection{The agile journey at Company $D$}
Company $D$ was described as a follower and had had a rigorous methodology mostly, with long projects of twelve to fourteen months or even longer time to delivery. These long lead times had often not satisfied the customer requirements or they had exceeded their budget. People in the organization had heard about agile project management and they had eventually hired a third-party consultant and had conducted a pilot with one enhancement project. The developers had paired with the vendor's developers to learn the practices and the behavior. At the point of this research, the organization had been on a lean journey for five to seven years and they connected that experience to the agile principles in a three-day course before they started their agile teams. They were not implementing vanilla Scrum, but they still had iterations, daily stand-ups, retrospectives, and measured velocity even if they did not have two-week sprints. They saw a challenge in transitioning from waterfall to agile since they still somewhat had the old culture of defining all the requirements before the development started. 

\paragraph{Interviewee at Company $D$} 
The interviewee from Company $D$ was part of the company's enterprise IT division and involved in new development and support work, but also worked with responsibilities in the delivery of business capabilities. The interviewee was on the development side and had been for a couple of years, and managed the project managers.

\subsubsection{The agile journey at Company $E$}
Company $E$ had had challenges with aligning the business requirements with the IT part of their company. They had had low transparency in their process and had built whatever was the most urgent at the time, i.e.\ little formalization and quite ad hoc. The agile journey started with one manager taking a Scrum Master class and trying out the Scrum framework in a pilot project with a team of four members. At the point of this research, they had applied Scrum or agile concepts to multiple areas but the longest was within IT portfolio management. They used a product backlog in form of a bulletin of stories available across functions with a designator of the department who acted as a product owner if a request was made. They also did sprint planning, but in two steps; the first being a review of the prioritized backlog and tracking of throughput, and then, in a second meeting, they had task planning in order to commit to a set of tasks for the coming three-week sprint. They, recently before the interview, went from having daily stand-ups with four people to 16 people invited, which increased the length from five or ten minutes to twenty minutes.  

\paragraph{Interviewee at Company $E$} 
The interviewee from Company $E$ was managing the team that was responsible for the project portfolio management process, which also meant that the interviewee was responsible for the agile transformation. The focus was on the large initiative but they had started with small enhancement production support work with a pilot. The introduction of agile to that team was by sending them to a two-day training, one day at an external organization to learn the framework, and one internal training workshop to describe how to apply the framework within their own organization.

\subsubsection{The agile journey at Company $F$}
The first interviewee was from a part of the organization where they were still in a pilot phase following a fixed list of requirements. Therefore, they only leveraged Scrum for their IT resource tasks. The second interviewee was from the IT organization at Company $F$ that had around 800 employees. The PMO organization had around 25 project managers that were traditionally focused on a waterfall methodology. Two years before this research was conducted, the CIO had heard about agile projects and had decided that the organization was to contact a consultancy firm in order to investigate the possibilities further. At the time of the interviews, they had agile start-up squads that initially worked side-by-side with the teams until the new agile teams were more or less independent. They had a high level of engagement in the beginning and sometimes also actually facilitated the stand-ups, the iteration planning, and the retrospectives until the Scrum Master was confident enough to take over. The biggest benefits they had seen so far were visibility and accountability and stated that they saw agile as more disciplined than waterfall because of these factors. At the point of this research they had a range of both traditional waterfall projects and agile dittos. Their traditional approach was not described as very strict and they called it ``free-fall" instead of waterfall, since they often did not have formal stages. The strictness depended on what the project manager wanted.

\paragraph{Interviewee 1 at Company $F$} 
The first interviewee from Company $F$ was the team leader of a team supporting order management, pricing, and configuration management for the sales and distribution area.

\paragraph{Interviewee 2 at Company $F$} 
The second interviewee from Company $F$ was a part of the PMO focused on UK and North America and responsible for the agile implementations.

\subsubsection{The agile journey at Company $G$}
Three different employees participated from Company $G$. The first one was from a context where they created application to aid the implementation of an ERP system. The reason why they had started with an agile approach was in order to have more customer contact during the development but also in order to develop and deliver faster. The team in this context was described as being different with regards to some aspects, mainly being that the developers were not collocated and not 100\% dedicated to the project. These aspects were described as being the cause of some challenges when they needed fast delivery. The Scrum framework was used to provide the intense communication needed to finish the projects faster, but also in order to get all stakeholders on the same page. Before using an agile approach, they had used a traditional waterfall process. Their agile framework included daily scrums, sprint planning, reviews, retrospectives, and a burn-down chart. Their requirements remained fairly stable over time during a project and did not change in unpredictable ways. 

The second interviewee worked with two different teams; one worked on service development and the other created mobile applications. They had one product owner for both teams who prioritized the backlog, and they had two-week sprints. They did sprint planning, review, daily meetings, and weekly meetings with the product owner and the teams. The teams had been formed as agile teams directly and therefore they had not used any other methodology before with these specific teams. The benefits that they did not think they would have had with a tradition process, were the proximity with the project owner and to get feedback from the customer continuously. The biggest challenge described at the point of this research, was the ``done'' criteria since it meant different things to different roles. Now they discussed this issue and re-defined ``done'' as finalized, tested, and reviewed. 

The third interviewee was in the global IT organization of the company and the teams created new solutions for implementation projects. The current project had started one year before this research was conducted and was described as different from the other projects since it was business-driven instead of driven technically. The current project was described as very innovative, which was also stated as the reason for having implemented an agile process. The Scrum or agile methodology had been undefined at the beginning of the project and a lot of experience in the previous year had been used to improve their current process. The focus on adopting continuous improvement was described as one of the most useful practices in place. They stated that they were implementing the Scrum framework well and were content with the process and with their agility in general. One challenge in the beginning had been to convince the business organization of the benefits of an agile approach, but at the point of the interview, they stood behind the process and were actively checking the backlog and also accepted the backlog as the leading artifact for projects.

\paragraph{Interviewee 1 at Company $G$} 
The first interviewee from Company $G$ was part of a relatively new organization within the company. They worked within business analytics in a group focused on deployment of new solutions. The interviewee was the Scrum Master\slash project manager (the lead of the development process) of one team at the time. The interviewee had been involved in five such projects using the Scrum framework. The interviewee was the one that introduced an agile development processes at the organization after taking a Scrum Master training and then educating the team.

\paragraph{Interviewee 2 at Company $G$} 
The second interviewee from Company $G$ was the Scrum Master of two teams. The interviewee had previous experience with with the Scrum framework, which was a requirement for the current employment. The teams had no external training but had been trained on the job by the interviewee.

\paragraph{Interviewee 3 at Company $G$} 
The third interviewee from Company $G$ was part of the global IT organization at the company and responsible for the development process. The interviewee had started by taking an internal agile and Scrum training course but had also later taken a Scrum Master training class. The interviewee was the Scrum Master of a team working on new solutions for application services, but half of the team was from IT and half was from the business side.

\subsubsection{The agile journey at Company $H$}
\emph{This company only participated in the survey so the following background is taken from a shorter, unrecorded, phone interview with the Scrum Master of the participating project and was written down and summarized afterward. The reason why the Scrum Master was not contacted again was that top management stopped the project as a part of a larger effort to save money, shortly after the survey was filled in.}

The project consisted of developing an existing enterprise system. The system was the first enterprise system project they had had in the organization. The software under development was a safety-critical system and the organization wanted to integrate it into the rest of the organization. The organization still used a traditional stage-gate waterfall system, and they all had to adapt to, and work against, that rigor. The gates were fixed and they had to calculate an end-cost for the entire project. The idea was then to work with agile methods in-between the gates. The business part of the project had been going on for half a year, and the project had two-week sprints with specifications in connection to these. They always met at a meeting on day five, where the requirements were written down and documented (user stories for requirements were not being used). They received requirements from their product owner within the organization who ultimately decided their priority. The also used prototypes and the first mockup had been shown to the customer very early in the project life-cycle.

\subsection{Interviews}
The interviews were semi-structured and set out to identify aspects of the agile transitions that might not have surfaced in the survey, as well as providing a deeper investigation of the group development and maturity in connection to these agile transitions. The questions were of comparative nature in order to investigate the differences between their earlier process and what changed when the agile approach was introduced in the company. The main questions asked were:

\begin{itemize}
\item What is the agile history of the organization and why did you choose to implement agile practices in the first place?
\item What methods were used before that? 
\item What agile methods do you use now and why?
\item What do you think is working\slash not working and why?
\item How much training was conducted in connection to changing to agile methods?
\item Do you see a difference in how high performing teams adopt agile compared to newer or less mature teams?
\item Do you think that agile methods affect group cohesion?
\item Do you see increased job satisfaction?
\item Are the teams collocated and, if not,  what are the challenges of geographically spread out group members?
\item How are the agile practices combined with the surrounding environment within the organization? 
\end{itemize}

The interviews were conducted over teleconference and recorded with the permission of the interviewee. The interviews were transcribed word for word and then a content analysis was conducted by the first author to find statements regarding the transition to an agile approach with regard to the presented aspects of group development theory. The content analysis involved marking statements under the categories that were related to building agile teams, and calculating how many interviewees that had mentioned the same aspect. A summary of the interview result is found in Section~\ref{sec:resultsint}.

\subsection{Surveys}\label{sec:methodologysurvey}
The surveys used in this study were the developer survey as suggested by \citet{sidkyphd} presented earlier in the new categories as suggested by \citet{grenjss} with the scale 4 part of the Group Development Questionnaire \citep{wheelan} added in the beginning (six factors\slash groups of items in total). The agility survey for developers and the GDQ scale 4 were put together in online surveys containing 31 items in total for group-members (from both collocated and distributed agile teams) to answer on a Likert scale from 1 to 5 (where 1 = low agreement, and 5 =  high agreement to the statement). It was the manager who selected which agile teams to have participate in the study, but we requested that they would select different types of teams with regards to performance\slash maturity. All the items from the agility measurement are presented below. 

There is an overlap between quantitative data from our earlier publications and this study. The preliminary results published in \citet{gren2015seaa} consisted of 45 employees from two of the companies also included in this study. A subset of the data was also used in the validation study published in 2015 \citep{grenjss}. This paper has an addition of 21 data points (47\% more) for the quantitative part. The analysis of data on factor-level in the correlation matrix is is also new in this study. 

In order to say which group stage a group is in, the whole 60-item GDQ survey must be used. However, it is possible to only measure the degree of effective group work by using scale four of GDQ. That scale was used to correlate the GDQ to other effectiveness measures, like e.g.\ patients' outcomes in emergency surgery teams \citep{wheelan20032}. This boils the number of items down to 15, which makes the time put into the study by participants minimal. Only three example questions are presented here due to copyright reasons, however, it includes the three facets ``Effective organization'' (how well the group organizes its work), ``Culture, norms, values'' (productive group norms, participatory and open culture, and values), and ``External relations'' (how the group integrates with the surrounding ecosystem).

\paragraph{Group Development example questions from Scale 4 (Work and productivity)}
\begin{itemize}
\item The group gets, gives, and uses feedback about its effectiveness and productivity.
\item The group acts on its decisions.
\item This group encourages high performance and quality work.
\end{itemize}

\paragraph{Agility Factor 1 (Dedication to Teamwork and Results)}
\begin{itemize}
\item You are willing to dedicate time after each iteration\slash release to review how the process could be improved.
\item You are willing to undergo a process change even if it requires some reworking of already completed work products.
\item Your team members seek your input on technical issues.
\item In a group meeting, the customer suggested something about the product. You disagree and have a better idea; it is acceptable for you to express disagreement with your customer and suggest something better.
\item Your manager seeks your input on technical issues.
\end{itemize}

\paragraph{Agility Factor 2 (Open Communication)}
\begin{itemize}
\item There should be a mechanism for persistent knowledge sharing between team members.
\item People should use a wiki or a blog for knowledge sharing.
\item When you run into technical problems, you usually ask your team members about the solution.
\item The organization values you and your expertise.
\end{itemize}

\paragraph{Agility Factor 3 (Agile Planning)}
\begin{itemize}
\item You usually participate in the planning process of the project you are working on.
\item You participate in the planning process of the project you will work on.
\end{itemize}

\paragraph{Agility Factor 4 (Leadership Style)}
\begin{itemize}
\item Your manager listens to your opinions regarding technical issues.
\item Your manager encourages you to be creative and does not dictate to you what to do exactly.
\item You do a better job when choosing your own task on a project instead of being assigned one by your manager.
\end{itemize}

\paragraph{Agility Factor 5 (Honest Feedback to Management)}
\begin{itemize}
\item If your manager said or did something wrong, it is acceptable for you to correct and\slash or constructively criticize him\slash her face to face.
\item It is acceptable for you to express disagreement with your manager(s) without fearing their retribution.
\end{itemize}

\subsection{Procedure}
Ten 30 to 45 minute open-ended interviews were conducted with a manager of seven out of eight participating companies with an overall perspective of their journey towards working with an agile approach. The main reason for interviewing managers was to get deeper qualitative data on both agility and group maturity in the organizations.

The surveys were sent out to the employees via email by their manager. The survey was created as an online questionnaire and the link to it was shared in the email. The responses were anonymous and not seen by the manager. The survey started with the GDQ4 questions followed by the agility survey for developers. It was sent to 109 employees in total of which 66 replied, i.e.\ a response rate of 61\%. This response rate is above average (52.7\%) within organizational research~\citep{responserate}. One reminder was sent via email by one of the managers (from one of the organizations). Filling out the survey took approximately ten minutes and all the questions were compulsory.

\subsection{Quantitative Data Analysis}
The first step to see if there is a connection between agility and group development was published in \citet{gren2015seaa}. The current study's analysis contains 47\% more data points and is divided into the factors found in \citet{grenjss}. However, with this relatively small sample the quantitative part of this study should only be seen as a supporting complement to the qualitative analysis, that does not seem to contradict our other findings.

GDQ is a thoroughly validated tool \citep{wheelan}, but the agility measurement is not. Based on the validation studies of GDQ \citep{wheelan} the effect size is considered high in this case. According to \citet{cohen}, a multiple correlation analysis with 6 variables needs a sample size of $N = 45$ for $\alpha$ = .05 and Power = .80. The agile factors from \citet{grenjss} were correlated to the overall GDQ Scale 4 mean values, and based on a sample size of $N = 66$, we would get a power of 99\% if we have a high effect size of .5. The main issue here is that we have relatively few data points from four different context (or companies), so the effect size could be expected to be lower, however, even with an effect size of .4 we would have a 92\% chance of finding an effect, if there is one (statistical power $= 1 - \beta$). For more details on prospective power analysis, see e.g.\ \citet{murphy2004spa}.

\begin{figure}
\centerline{\includegraphics[scale=0.5]{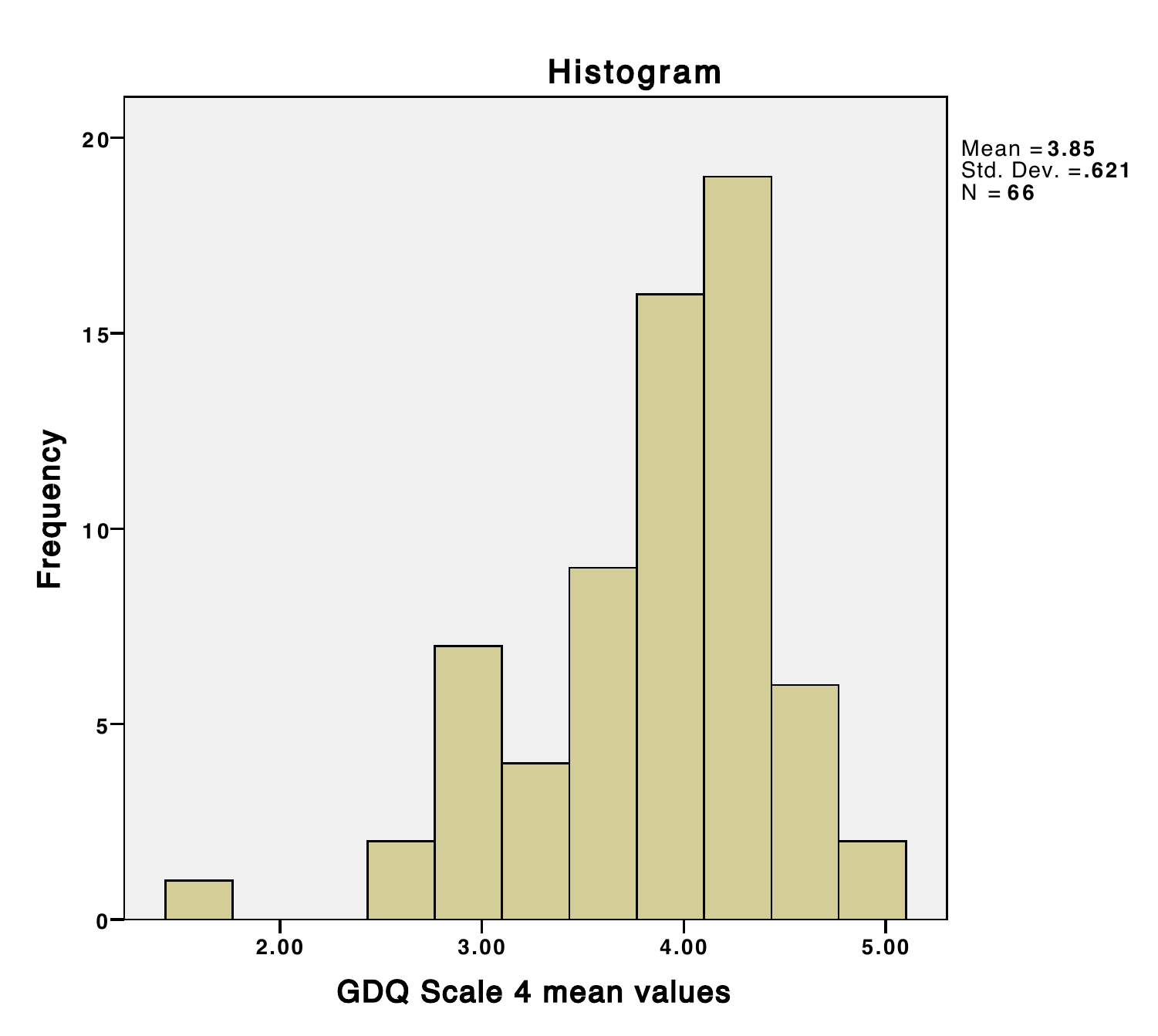}}
\caption{Frequency histogram for the factor GDQ Scale 4 mean values.}
\label{d}
\end{figure}

\begin{figure}
\centerline{\includegraphics[scale=0.5]{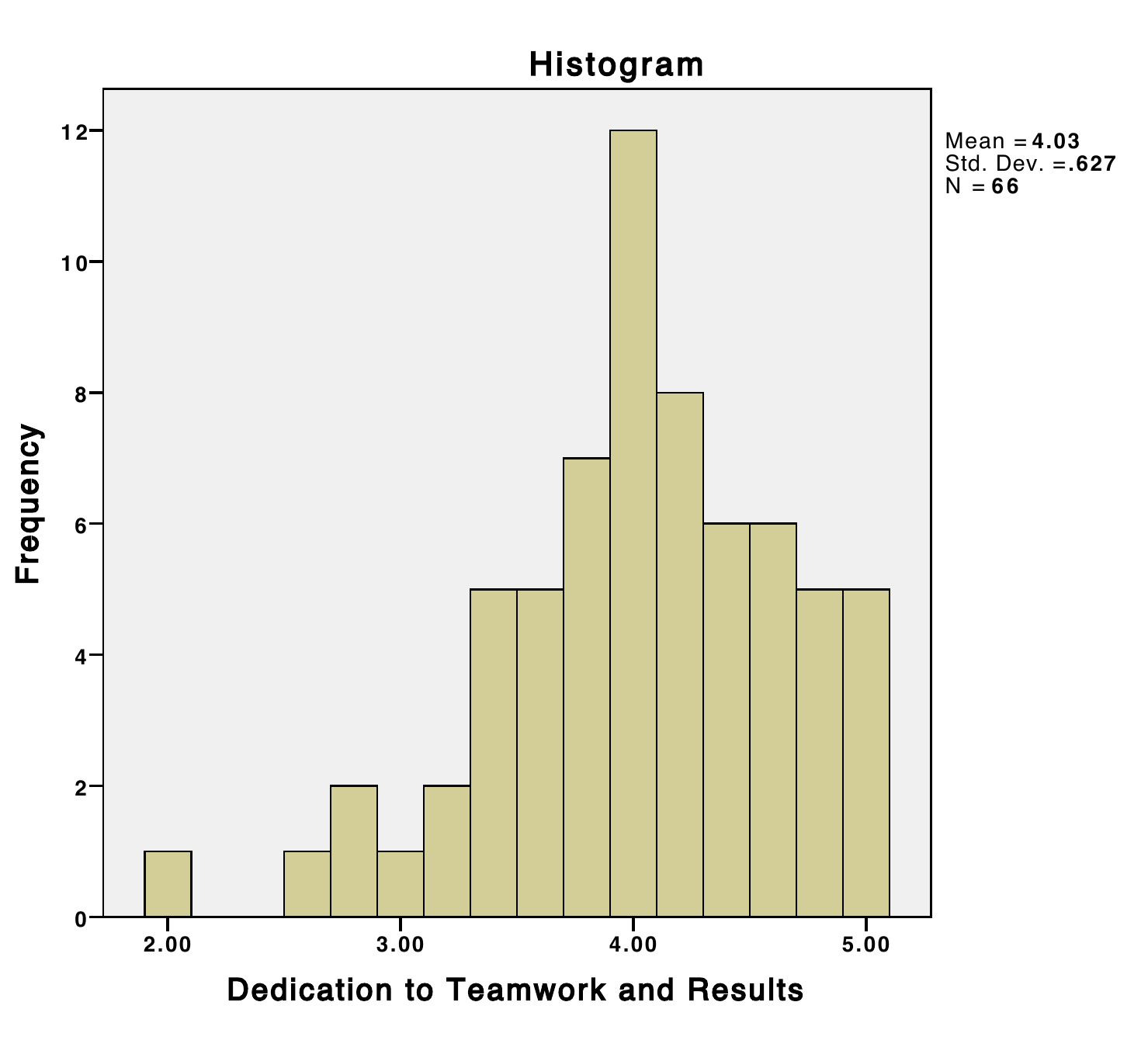}}
\caption{Frequency histogram for the factor Dedication to Teamwork and Results.}
\label{d2}
\end{figure}

\begin{figure}
\centerline{\includegraphics[scale=0.5]{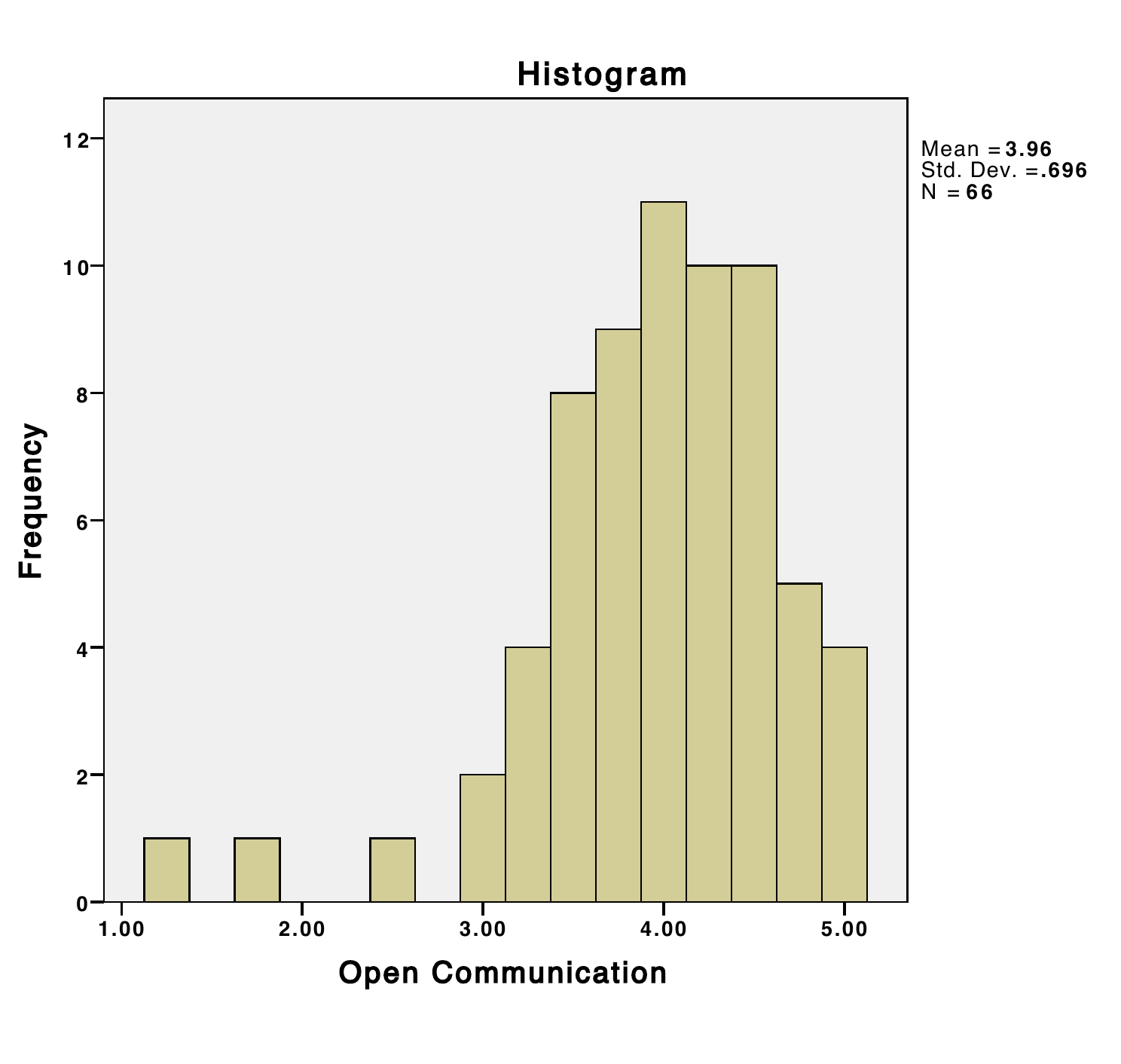}}
\caption{Frequency histogram for the factor Open Communication.}
\label{d3}
\end{figure}

\begin{figure}
\centerline{\includegraphics[scale=0.5]{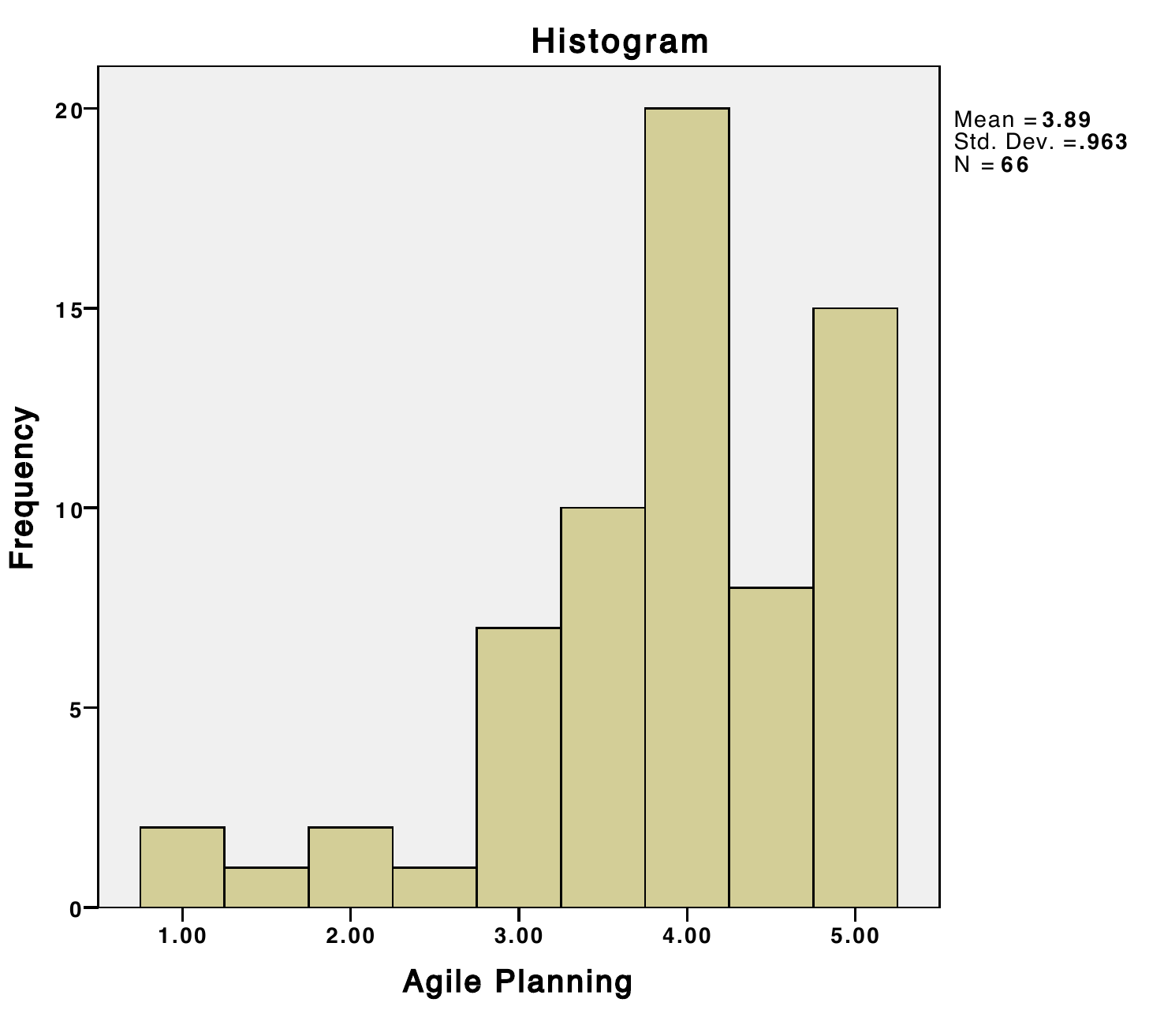}}
\caption{Frequency histogram for the factor Agile Planning.}
\label{d4}
\end{figure}

\begin{figure}
\centerline{\includegraphics[scale=0.5]{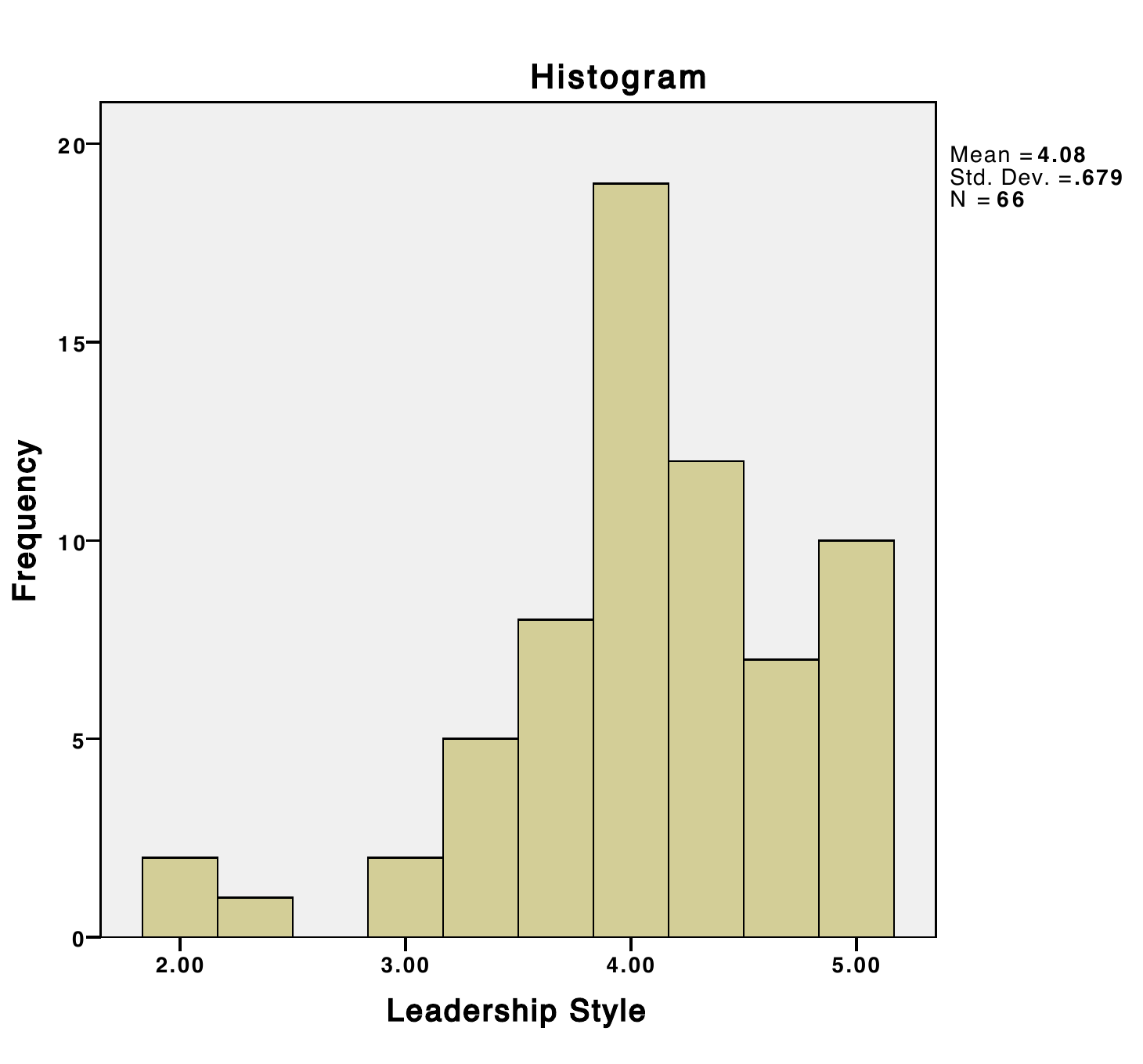}}
\caption{Frequency histogram for the factor Leadership Style.}
\label{d5}
\end{figure}

\begin{figure}
\centerline{\includegraphics[scale=0.5]{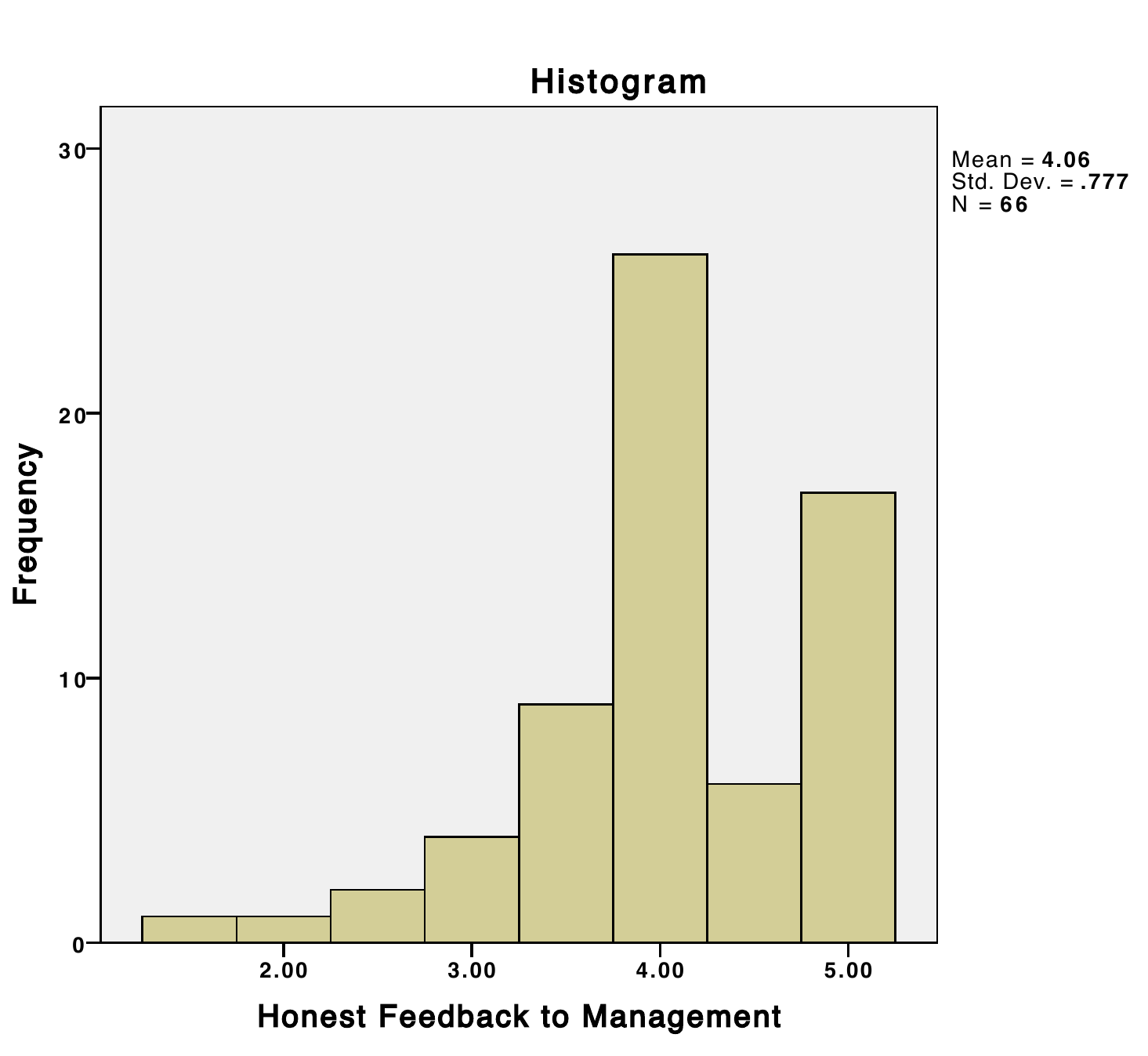}}
\caption{Frequency histogram for the factor Honest Feedback to Management.}
\label{d6}
\end{figure}

In order to evaluate if the data was normally distributed we plotted frequency diagrams for all the six factors (see Figure~\ref{d}, Figure~\ref{d2}, Figure~\ref{d3}, Figure~\ref{d4}, Figure~\ref{d5}, and Figure~\ref{d6}). We saw some concerns with skewed data and ran the Shapiro-Wilk test for normality (see Table~\ref{shapiro}), which were significant for all factors, i.e.\ we had an issue with the normality assumption. Therefore, we chose to use Spearman's $\rho$ instead of Pearson's $r$ in our correlation analysis, since it is based on ranks instead and is therefore nonparametric, i.e.\ does not assume any distribution. Spearman's $\rho$ also allows, and compensates for, tied ranks.

\begin{table*}
\renewcommand{\arraystretch}{2}
\caption{Shapiro-Wilk Tests of Normality ($N=66$).}
\label{shapiro}
\centering
\begin{tabular}{cccc||c||c||c||c||c}
\hline
\bfseries  & Shapiro-Wilk Test Statistic \bfseries & Sig.  \\
\hline\hline
GDQ Scale 4 mean values & .935 & .002 \\
\hline
Dedication to Teamwork and Results & .957 & .023  \\
\hline
Open Communication & .900 & .000 \\
\hline
Agile Planning  & .878 & .000 \\ 
\hline
Leadership Style   & .906 & .000  \\
\hline
Honest Feedback to Management  & .879   & .000 \\ 
\hline
\end{tabular}
\end{table*}

\section{Results and Analysis}\label{sec:results}

\subsection[Summary of Interviews\ldots]{Summary of the Results from the Interviews}\label{sec:resultsint}
The following themes emerged from the open-ended questions in the interviews and in the analysis below we connect these themes to group development theory. 

\begin{itemize}
\item Increase in job satisfaction
\item Situational leadership
\item Collocation issues
\item The discipline of agile project management
\item Teams already agile before implementing agile practices
\item Explicit group developmental aspects
\item Personality 
\end{itemize}

The results are first summarized and then exemplified by one or a couple of quotes. When an interpretation or analysis is provided in connection to group development or maturity this text is written in \emph{italics}.

\paragraph{Increase in job satisfaction}
Since work satisfaction has been shown to correlate to higher values of GDQ4, and there are some studies showing that job satisfaction is higher on agile teams (see Section~\ref{intro}), we asked all the interviewees if they see an increase in job satisfaction with the implementation of agile practices and values in connection to building their agile teams. All of the ten interviewees said that they had seen, or experienced, higher job satisfaction since they changed to an agile approach. When asked why they believe the job satisfaction increased, four interviewees stated the empowerment of the team members as the main reason. These interviewees used terms like ``feeling of ownership,'' ``influence,'' ``involvement in decisions,'' ``seeing the progress,'' ``visibility'' and ``autonomy.''

\begin{quote} 
``Our developers, you know, giving them the autonomy and let them be self-directing and responsibility and authority so they can influence what they do and how they do it, and also how they work together, those folks who have been working on these teams have a much higher work satisfaction, then they did in their developer roles before.'' [Team Lead\slash Scrum Master]
\end{quote}
\emph{For all group members to engage on such high level, the group probably needs to be at the later stages of group development. }

Seven interviewees mentioned the team spirit as being the main reason for the increase in job satisfaction (one interviewee mentioned both empowerment and team spirit). One interviewee specifically mentioned the effects of having cross-functional teams with the business side represented in the team. The ``us versus them'' mentality had therefore disappeared. Two other interviewees stressed how teams focusing on being collaborative and self-organized show higher job satisfaction than the teams that do not. Two interviewees also made the explicit connection themselves of how the agile approach enables and gets the team to get together and create team spirit. 

\begin{quote} 
``Yeah, we've actually had some discussions about that: people just think it's more fun to be on an agile team. They enjoy being on what I call a close-knit team, more than if they are just off doing their tasks without much group interaction. So yeah we have seen that, well they just describe it as having more fun. I think that's a collection of the relationships and the collaboration and the sense of team when they're together as an agile project or team.'' [Agile responsible at a PMO]
\end{quote}

\begin{quote} 
``Yes, I believe simply because it's more satisfying for people to work in such a team that maybe then also the team development is easier or faster.'' [Scrum Master at a global IT organization]
\end{quote}

Two interviewees also mentioned stress as a factor that is different as compared to more traditional project management. However, one interviewee stated that stress was lower due to the stable work rhythm and the concept of the manageable increment and not worrying about future deliveries. The other interviewee mentioned an increased stress with agile because of prototyping. However, that same interviewee later mentioned that their product owner was unavailable and brought changes to the review meetings instead of earlier when the team welcomed them, which would indicate that this was due to the product owner not acting as expected in Scrum.  

\paragraph{Situational leadership}
All interviewees were very engaged in their teams and seemed to have reflected to a large extent on what the teams might need in order to increase their effectiveness and well-being. \emph{One key in agile processes is the self-organization of teams, but group psychology shows us that teams need leadership and also different kind of leadership depending on the group development stage.} Two interviewees explicitly described some of their behavior as ``non-agile'' and apparently felt a need to explain that they saw a need from the team to be managed at some points in time (especially in the beginning of projects).

 \begin{quote} 
``I don't know if I go beyond agile, but I think I probably go beyond the typical Scrum Master role. I really try to mentor the team and I try to help them work out conflicts, and if I see that there's some risk somewhere, I call them on it. I sort of take on a team lead role. I don't know if that's pure Scrum Master, I'm a project manager and developer team lead. /\ldots/ I'm not the technical lead, but I do manage them in the sense of more traditional project management.'' [Scrum Master\slash Manager]
\end{quote}
\emph{From a group development perspective (as presented in the Section~\ref{sec:related_work}), helping the group with conflict management is key to progress in stage two of the group development. Only describing the desired end-state of agile high performing teams seems to have gained foothold with some practitioners and the road to getting there (i.e.\ the great need of good situational leadership) seems to be lacking.} Another interviewee with more overall responsibility of different agile teams also highlighted the importance of good leadership in order to move the teams forward. 

 \begin{quote} 
``At least it's very critical, at least from what I've seen, to have some leaders on the team that can help kind of drive and push and continue to elevate the team as the cycle moves along.'' [Responsible for (and initiator of) the agile transition]
\end{quote}

Another interviewee described a bad experience with task volunteering since all seven group members worked on the same task and did not take responsibility for when it would be finished. Therefore, the interviewee stated that there is no possibility for task volunteering at that point in time. \emph{From a group developmental perspective, group members will not volunteer for tasks to a large extent when the group is less mature. Newly formed work-groups in fact benefit from more directive leadership in the beginning, which should not be seen as a failure.}

\paragraph{Collocation issues}
All interviewees replied that their agile teams would have benefited a lot from being collocated, if they were not. All of them also explained how clearly they see a difference when they are collocated and then spread out. What the interviewees all said they do to bridge the gap is very much in line with the research conducted by \citet{noll2010global} and presented in Section~\ref{intro} (i.e.\ site visits, synchronous communication technology, and knowledge sharing infrastructure). However, when one interviewee compared the new agile team-based organization with a mix of collocated and distributed teams, they see a clear difference in the success of building these teams.

\begin{quote} 
``In IT for a while even before agile the IT people didn't have assigned desks or space. They have a universal hotelling set up. People were not used to go to one desk every day. When agile puts people in the same room, they got some space to be proud of. Now they got a place to go to and count on it. That structure was beneficial. The distributed teams continued within the non-belonging. The camaraderie or team identity, and hold each other accounted, are still problems within the distributed teams.'' [Enterprise agile coach]
\end{quote}
Four of the interviewees only explained the vast difference and that collocation is utterly important and tried to have as many site visits as possible, especially in the beginning with newly formed teams. However, two interviewees saw it a bit differently. Since it is often a matter of resources having to be distributed (as mentioned by four of the interviewees), six of the interviewees were trying to create their own virtual process that would build as much team spirit as possible \emph{(i.e.\ move the group through their development)}. One interviewee even state that having an agile process with distributed teams is a must in order to even have a chance at building a team at all.  

\begin{quote} 
``I think that, again there are people within [the company], and I'm sure everywhere, who feel that you can't do real agile if your team is not collocated, and if the members aren't dedicated. Not only do I disagree with that but I also think that it really helps in a distributed atmosphere /\ldots/. The way you get these people together in these intensive meetings everyday. I think it really speeds up the team coalescing, the storming phase pretty early on to work stuff out.'' [Scrum Master\slash project manager]
\end{quote}
One had an idea of using video in addition to audio, and another one had created team building ceremonies online with open discussion meetings where everyone dialed in, filled out backlog items together, and had a morning routine of looking at the new burn-down chart together online. \emph{Making use of modern technology in such a way is of course a very good idea from a group developmental perspective. Having a strong leader that sets such a psychological contract of the process and who does not skip such practices even in the storming phase of Stage two, is of course a key.} One interviewee had a more traditional work-group transitioning to an agile process and explained the great advantages of having daily stand-ups, since remote resources actually felt more as a part of the team with such a practice in place. 

When discussing collocation one interviewee strongly underlined the fact that collocation must be in the same room having the desks like an enclave so that one can lean over and communicate in a second. Having people in different rooms was stated as as bad as having resources in different countries.

\begin{quote} 
``I just think that, if you are collocated your communication level just goes up exponentially. I'm even talking about, you know, feet versus yards versus different floors. If you can be within feet of each other it's so much better. Collaboratively and as far as levels of communications, but as soon as you're on a separate floor, well I think the difference between a separate floor and [a country on another continent but the same timezone] is about the same distance. If that makes sense.'' [Agile responsible at a PMO]
\end{quote}

\paragraph{The discipline of agile project management}
Three of the interviewees spoke about that they had heard that agile is more laissez-faire and less controlled than more traditional approaches to projects. However, their practical experience with agile project management was exactly the opposite, i.e.\ more disciplined because of the even work rhythm and daily ceremonies. 

\begin{quote} 
``The one disadvantage, I don't even know if this is a disadvantage, but there are some temperaments that don't lend themselves well to the kind of discipline that this requires in terms of showing up to a meeting on time. Inevitably, there will be one team member that just can't get there on time. And that's frustrating. So I would say the structure is not flexible. It's not meant to flexible, and there are some people that just cannot deal with that.'' [Scrum Master\slash project manager]
\end{quote}
They all mentioned the fact that when transitioning to an agile process, some issues surfaced that were invisible before, such as overallocation, over-staffing, projects stalling, and people not delivering on time.

\begin{quote} 
``That [responsiveness to change] is probably the biggest [advantage], and the visibility and accountability. Some people would think that agile is less disciplined, but in reality I think agile is more disciplined. Because you create that visibility and that accountability on a daily basis rather than, like in a typical waterfall project a lot of times there is a check-in a week later /\ldots/. And it's a little bit out of site, out of mind. Whereas the agile methodology keeps it in the forefront, and you really can't say things like: `I didn't look at that.' too many days in a row without it standing out. I think one of the things we kind of say is that agile doesn't like low performers because they stand out. So that's one of the benefits: that you can recognize that somebody is not performing well, because things are being delayed delayed delayed, or during the daily stand-ups you quickly see that they are either not focused on the project or the work. Or you might have a skill gap or something like that. /\ldots/ they stick out like a soar thumb. And what happens is that they can drag the whole team down because it can be infectious, so to speak.'' [Agile responsible at a PMO]
\end{quote}
\emph{From a group development point of view, all team members need to contribute and be needed for the group-goal fulfillment, which means that group membership is at least as important as group leadership for the group to develop to a high performing team.}

\paragraph{Teams already agile before implementing agile practices}
Four of the interviewees mentioned that the transition to an agile approach was very easy for some teams, since they were already working in such manner even before the introduction of agile practices. \emph{which, from an agile consultancy point of view (the view that agile project management is something new), would be impossible. However, this also supports the idea that the concept of an ``agile team'' cannot be entirely new and connecting it to group maturity would partially explain why practitioners hear, and experience, so many success stories of such teams.} 

\begin{quote} 
``So this ever-adapting mindset of agile, well, this team is very suitable for that and very receptive to that approach. It has a lot of creative and very engaged people on it, they love that they have the freedom to change things and improve where it makes sense. So this is why this team grew through agile.'' [Agile project manager]
\end{quote}

\begin{quote} 
``It has been an easier adjustment if the team was more or less working that way anyway.'' [Overall responsible for the agile transition]
\end{quote}
Another interviewee also stated that the high performing teams adopted agile principles and practices very easily, and that it just provided some structure and where to focus from a priority perspective. 

\paragraph{Explicit group developmental aspects}
Nine of the interviewees explicitly made connection between group development and building agile teams, as opposed to traditional teams. One interviewee said that, with the agile practices, new members get integrated into the teams much faster and quickly becomes a part of the ``family'' and that people on the team like to work with each other. \emph{Such interpersonal attraction is a characteristic of a stage four group.} Lifting conflicts is also something that is enforced on the agile teams according to another interviewee, something that might be more difficult for some developers who are not used to such openness.

\begin{quote} 
``The challenge is maybe just changing the culture. [Interruption] it's definitely different. People are uncomfortable talking about the blockers that hinders the work being done for the week. They aren't used to speaking about those issues.'' [Agile team leader]
\end{quote}
One company also implement indefinite agile teams that take on different products and not projects, \emph{which means they have understood that once you succeed in building a high performing team, the best you can do is challenge them with new projects and complex tasks.}

\begin{quote} 
``I get the agile team up and running as an agile team and get them executed in a high capacity manner. They are basically centered on the enhancement and improvements of that product and basically the team is an agile team indefinitely, which means it never ends. Whereas in our case with projects, we will get a project going and get the momentum going and just about that time, they are really humming despond because the project was done and they reached the benefits.''  [Agile responsible at a PMO]
\end{quote}
\emph{When it comes to the causality discussion of if agile practices leads to group maturity or if group maturity leads to agility, this boils, again, down to the definition of agility. As mentioned in the Section~\ref{sec:related_work}, there is a known difference between ``doing agile'' and ``being agile'' as well as there is a difference between the agile principles and the agile practices. If we by ``being agile'' mean the agile principles and the cultural change they imply, we believe we see support in our interview data for an overlap, or an interaction effect, between what is meant to be an ``agile team'' and a mature group from a social psychology perspective.} However, in practice, one interviewee stated that high performers can double their performance on an agile team compared to being on a traditional project, but just because it is labeled as an agile team and doing agile practices does not mean it is a high performing team, that depends on the group itself (i.e.\ the people in it). Another interviewee also stated that as far as performance goes, it is more about taking on roles the group members usually do not take on in other teams, when joining one of the agile ones, and contributing more to the team's effort and not the individuals'. \emph{This also highlights the group maturity focus implied when transitioning to an agile approach.} Four explicitly stated that transitioning to agile teams is closely tied to developing the team from a ``group dynamics'' perspective, meaning they define the main challenge as the behavioral one (caused by continuously having to show results, communicate, discuss, negotiate, always getting feedback from the customer, taking responsibility, and being more reactive).  

\begin{quote} 
``So it's more about the relationship than the process, and making sure we are more clear about the roles and responsibilities.'' [Responsible for (and initiator of) the agile transition]
\end{quote}

\paragraph{Personality}
As mentioned in the Section~\ref{intro}, personality has been a popular starting point when investigating psychological aspects of software engineering teams. Five of the interviewees in this study also highlighted the importance of hiring the right people and the issue of skill gaps. The same interviewees also mentioned the fact that, with the high intensity of agile, that type of work is probably not for everyone. People who are engage and get empowered will flourish, but people who like to be given more direction in their work will not really be happy on an agile team, according to two of the interviewees. One interviewee also stated that high performing individuals do not make good Scrum team members since they often do not bring the right approach and behavior needed to be a team player. \emph{Our data also confirms that both team working skills, as well as individual skills are desired when on an agile team, just like a high performing one is described in the group development theory.} 

\begin{quote} 
``I think for some people they have both skills, but for some it's very clear, especially in software development, where it's clear that some are excellent at their role, but are not actually in a team, if that makes sense.'' [Responsible for (and initiator of) the agile transition]
\end{quote}
When it comes to the dynamic (instead of static) view of personalities, when developing a group into a team, people and their personalities can change, which we also found support for. \emph{In the group development theory presented in the Section~\ref{sec:related_work}, there is evidence showing that people do change their personality traits and their behavior depending on their context at a given time.} One interviewee had an experience of strong introverts actually taking more space on their agile teams and surprised the managers with regards to responsibility and drive. However, this interviewee also agreed with the other five stating that sometimes, a resource is not possible to integrate into the team at that point in time, but it is important to not make the fundamental attribution error, but instead also look at the contextual factors.

\begin{quote} 
``What I hear from my line manager who has several people here in the team also very interesting for me, as a feedback he says that he's quite surprised by how active the guys become in this team. So what you said about before\slash after. So it's probably as you said, they are interdependent. Agile requires that type of teamwork. But to a certain extent it also motivates for it, it's what I would say.'' [Scrum Master at a global IT organization]
\end{quote}
\emph{We think this shows that agile project management is not the silver bullet it is sometimes portrayed as in the agile literature. A major part of it is building high performing teams, which is, and has always been, difficult.}

\subsection{Correlation between GDQ4 and the agile measurement}

As can be seen in Table~\ref{fig:corrmatrixapa}, all agility factors were significantly correlated to the group maturity measurement. This implies that the participants' high and low scores on each item followed each other in a positive trend, i.e.\ a person scoring high on any of the agile factors, also gave high scores to the GDQ Scale 4 items. In order to clarify the results we need to elaborate and discuss the correlations in connection to the interview results. This is done in the next section (Section~\ref{sec:discussion}).

\begin{landscape}

\begin{table}[ht]
\begin{center}
\small
\caption{Correlation matrix for all the agile factors and the GDQ Scale 4 mean values.}
\label{fig:corrmatrixapa}
\centering
\begin{tabular}{LS@{}S@{}SLSLSLSLSLS}
\toprule
\multicolumn{9}{l}{Spearman's $\rho$ Correlations ($N=66$)}  \\
\midrule
\multicolumn{1}{@{\hspace{1em}}l}{Measure} & \mc{A} & \mc{B} & \mc{C} && \mc{D} && \mc{E} && \mc{F} \\
\midrule
A. GDQ Scale 4 (group maturity)  &  1 &  .416**  & .462** && .473**  && .277* &  & .265*    \\
B. Dedication to Teamwork and Results (agility) &   .416**  &  1 & .305*  && .400** &  &.445**  &&.501** \\
C. Open Communication (agility)&    .462**  &  .305* & 1  && .398** &  &.391**  &&.388**   \\
D. Agile Planning (agility) &    .473**  &  .400** & .398**  && 1 &  & .436**  && .333**   \\
E. Leadership Style (agility) &    .277*  &  .445** & .391**  & &.436** &  & 1  && .317**   \\
F. Honest Feedback to Management (agility)  &    .265*  &  .501** & .388**  && .333** &  & .317** && 1   \\

\midrule\\[-2.5ex]
\multicolumn{9}{l}{*p$<$.05, **p$<$.01}  \\
\end{tabular}
\end{center}
\end{table}

\end{landscape}

\section{Discussion}\label{sec:discussion}
The interview results show a large overlap between how agile teams are described by practitioners and how high performing teams are described in social psychology, and we therefore have answered our research question of how group maturity is connected to building agile teams. The result shows that the people responsible for the agile process (coaches, Scrum Masters or managers) all identify group issues as key success factors in building their agile teams. This might be evident to some, but the extensive work on a psychological level when building agile teams is not recognized in the description of the agile frameworks as much as seems to be needed, i.e.\ they are key success factors implicitly worked on by practitioners. The interviewees' comparisons of their previous methods and the transition to an agile approach indicate that such a journey cannot succeed without the group development being in focus and worked on extensively. This means that work-groups trying to become more agile might be helped by adding the group psychology perspective to understand what they need in addition to the concrete process they aim at implementing, however, this remains to be explored. It might at least help agile teams to further understand the connections between the agile principles and the agile practices and what the differences are in practice.

The correlation analysis also support the connections between the concepts and imply that the agile factors try to pinpoint much of what is captured in the group maturity measurement. The highest correlations between agility and group development were between:

\begin{itemize}
\item GDQ4 correlated to ``Dedication to Teamwork and Results'' ($\rho=.416$) and to ``Open Communication'' ($\rho=.462$) and to ``Agile Planning'' ($\rho=.473$).
\end{itemize}
This means that, if we assume that the agile factors measure some aspects of agility, a more mature group as defined in group psychology is also a more agile group, which confirms the results in \citet{gren2015seaa}. However, the current study provides an understanding of how practitioners work on these psychological aspects and their reasoning of the connection between agility and group maturity. Since the group developmental aspects are somewhat worked on by practitioners ``under the hood,'' explicitly extending the agile models with that of group development might help agile implementations to succeed better and faster. The fact that the agile factors are correlated internally as well, indicates and provides support for their own internal consistency (a reliability aspect in validation of measurements). 

In group development theory a group needs, as a mean value, 6 months to become high performing~\citep{wheelan2009}. Groups that have met less than 6 months are more unlikely to be high performing and therefore this study shows that they might be less agile in their work methods as well. 

We would also like to highlight the fact that, in this study, both the group development and the agility measurements were self-reported and on individual level, and therefore the connection is only between the individuals' perceived effective group work and perceived agility level. As presented in Section~\ref{sec:related_work}, the notion of the importance of the team in software engineering is not new \citep{weinberg}, nor is the need to adapt the team coaching to group readiness \citep{adkins}. However, these are anecdotal stories in popular books that do not provide a scientific approach to finding support for such claims. We believe this study provides some scientific evidence of the importance of the psychological aspects of group development when building agile teams. 

Another issue is that behavioral software engineering \citep{lenberg2015} aspects is often not in the curriculum of computer science education \citep{seovercoming}. This means that most technical staff lack the tools and approaches to deal with the psychological aspects of building teams. In the interviews, it was clear that the successful agile coaches must have obtained such knowledge either by having a different educational background or through reading such literature on their own.

We would also like to mention the lower correlations between between the agile measurements of ``Leadership Style'' and ``Honest Feedback to Management'' and GDQ Scale 4. We believe those management aspects are poor measurements of agility since they are taken from Sidky's~(\citeyear{sidkyphd}) way of measuring agile potential, and not existing agility. Surely, a manager can stop the team from becoming agile through directive and controlling leadership, but that is not a good measurement of what characteristics an agile and mature group has. In addition, the leader needs to act differently depending on the group's development stage (but also follower readiness as presented by \citet{hersey}), which means that the function of a Scrum Master (a Scrum facilitator that only guides the process and facilitates instead of directing) will be very difficult in a newly formed group. This paradox makes the agile practice of self-organizing teams something not to strive for if the group is immature (then the group development will be faster with clearer and more directive leadership). At these first stages, the group needs aspects of safety, inclusion, order, and structure, which means the group will not accept ``agile'' leadership. Only after the group has created a unified group culture and structure, will it be possible for the manager (Scrum Master) to withdraw and let the work-group be self-organizing. 

All in all, the two concepts (i.e.\ measurements), one taken from social psychology and the other from software engineering, are connected and we have an overlap with what we mean by an agile work-group and how a mature group is defined in social psychology. As mentioned before, this could be seen as a circular argument since effective team characteristics should also be true for effective agile teams, however, the details regarding how more agility implies more group maturity have not been research previously. It might be obvious that the apple falls from the tree to the ground, but having a model to explain the acceleration is still very useful. We think the link to group developmental psychology could provide useful guidance and some predictability to understanding team agility, and since ``agility'' is an undefined construct in both academia and industry, defining parts of we mean by an ``agile team'' as ``building a high performing team'' from a psychological perspective would be a useful definition and make the notion of an ``agile team'' easier to grasp and understand.

\subsection{Validity Threats}
The qualitative part of this study made it possible to assess how the agile practitioners work on group developmental issues in their agile process with higher internal validity, i.e.\ we could ask about the connections and causality from the perspective of the interviewee. The reliability of the qualitative interviews is, of course, lower since it is difficult to replicated such a design in detail. However, the survey used provide higher reliability since the exact same questionnaire can be distributed again. One of the largest threat in the quantitative part is the agility measurement, and since that tool is not thoroughly validated we cannot say with certainty that it measures aspects of agility (i.e.\ the agility measurement has issues with both content and construct validity). Furthermore, we have a too small sample (66 participants from four different organizations) in the survey to statistically assess the validity of the agility measurement. Therefore, the quantitative part of this study should only be seen as a complement to the interview study that, at least, do not contradict the qualitative result. Both the qualitative and quantitative data have, therefore, low external validity and we cannot generalize to a larger population of IT work group members. In addition, we only looked at the individuals' perception of agility and group maturity, which makes generalization to groups something we should do with care. This exploratory study should be seen as a descriptive first step in understanding how group maturity and the concept of an ``agile team'' are connected.

\section{Conclusions and Future Work}\label{sec:conclusions}
This paper set out to see how building agile teams is connected to group maturity. Through qualitative data from interviews and quantitative ditto from a survey, we have found that an agile team has many similarities to a mature group. This could increase the understanding of agility and partly help define an ``agile team'' since group maturity actually is one of the dimensions of agility. These findings are important contributions to both industry and academia since they might provide useful guidance and some predictability to understanding team agility. This study is descriptive in its nature since we wanted to try to explain the work being conducted implicitly by practitioners on the psychological dimension of building agile teams. 

In future research, also measuring the other group development stages by their corresponding scales would provide an understanding of how ``agility'' is connected to other types of challenges that groups go through in their more immature stages. Adding other dependent variables like, for instance, code quality and\slash or team productivity in further studies, would make an investigation on the interaction effect between agility and group maturity on such variables possible. Future work could develop more specific guidelines for how software development teams at different maturity levels adopt agile principles and practices differently. We also think studies with larger samples, both on individual perception, but more importantly on group-level would provide further understanding of how the psychological dimension integrates into different aspects of agility.

It would also be interesting to investigate if there is a correlation between the concrete agile practices and the group's maturity. We suggest that if such a study is to be conducted, or equivalent, the tool created by \citet{so} would be useful since it is validated on a larger sample ($N=227$) and include such concrete agile practices as a set of survey items. It would also be fascinating to analyze the verbal communication connected to group development in agile teams and compare them to non-agile ones by using Sequential Analysis of Verbal Interaction (S.A.V.I.) for example \citep{simon} or to data-mine developers' chat logs in their virtual work-place.

\section*{Acknowledgements}
This study was conducted jointly with SAP AG (http://www.sap.com) and we would like to thank internal SAP teams, the SAP customers, and Volvo Logistics, who were all willing to share information.

\bibliographystyle{model5-names}
\bibliography{references}

\begin{thebibliography}{59}
\expandafter\ifx\csname natexlab\endcsname\relax\def\natexlab#1{#1}\fi
\providecommand{\bibinfo}[2]{#2}
\ifx\xfnm\relax \def\xfnm[#1]{\unskip,\space#1}\fi
\bibitem[{Adkins(2010)}]{adkins}
\bibinfo{author}{Adkins, L.} (\bibinfo{year}{2010}).
\newblock {\it \bibinfo{title}{Coaching agile teams: {A} companion for
  ScrumMasters, Agile coaches, and Project managers in transition}\/}.
\newblock \bibinfo{address}{Boston, M.A.}: \bibinfo{publisher}{Pearson
  Education}.
\bibitem[{Balijepally et~al.(2006)Balijepally, Mahapatra \& Nerur}]{bali}
\bibinfo{author}{Balijepally, V.}, \bibinfo{author}{Mahapatra, R.}, \&
  \bibinfo{author}{Nerur, S.} (\bibinfo{year}{2006}).
\newblock \bibinfo{title}{Assessing personality profiles of software developers
  in agile development teams}.
\newblock {\it \bibinfo{journal}{Communications of the Association for
  Information Systems}\/},  {\it \bibinfo{volume}{18}\/}, \bibinfo{pages}{4}.
\bibitem[{Baruch \& Holtom(2008)}]{responserate}
\bibinfo{author}{Baruch, Y.}, \& \bibinfo{author}{Holtom, B.~C.}
  (\bibinfo{year}{2008}).
\newblock \bibinfo{title}{Survey response rate levels and trends in
  organizational research}.
\newblock {\it \bibinfo{journal}{Human Relations}\/},  {\it
  \bibinfo{volume}{61}\/}, \bibinfo{pages}{1139--1160}.
\bibitem[{Bion(1992)}]{bion}
\bibinfo{author}{Bion, W.} (\bibinfo{year}{1992}).
\newblock {\it \bibinfo{title}{Experiences in groups: {A}nd other papers}\/}.
\newblock \bibinfo{address}{London}: \bibinfo{publisher}{Routledge}.
\bibitem[{Boehm \& Turner(2003)}]{boehm20}
\bibinfo{author}{Boehm, B.}, \& \bibinfo{author}{Turner, R.}
  (\bibinfo{year}{2003}).
\newblock {\it \bibinfo{title}{Balancing agility and discipline: A guide for
  the perplexed}\/}.
\newblock \bibinfo{address}{Boston}: \bibinfo{publisher}{Addison-Wesley}.
\bibitem[{Boehm \& Turner(2005)}]{boehmm}
\bibinfo{author}{Boehm, B.}, \& \bibinfo{author}{Turner, R.}
  (\bibinfo{year}{2005}).
\newblock \bibinfo{title}{Management challenges to implementing agile processes
  in traditional development organizations}.
\newblock {\it \bibinfo{journal}{IEEE Software}\/},  {\it
  \bibinfo{volume}{22}\/}, \bibinfo{pages}{30--39}.
\bibitem[{Chronis \& Gren(2016)}]{kostasxp}
\bibinfo{author}{Chronis, K.}, \& \bibinfo{author}{Gren, L.}
  (\bibinfo{year}{2016}).
\newblock \bibinfo{title}{Agility measurements mismatch: {A} validation study
  on three agile team assessments in software engineering}.
\newblock In {\it \bibinfo{booktitle}{International Conference on Agile
  Software Development}\/} (pp. \bibinfo{pages}{16--27}).
\newblock \bibinfo{publisher}{Springer International Publishing}.
\bibitem[{Cohen(1992)}]{cohen}
\bibinfo{author}{Cohen, J.} (\bibinfo{year}{1992}).
\newblock \bibinfo{title}{Quantitative methods in psychology -- {A} power
  primer}.
\newblock {\it \bibinfo{journal}{Psychological Bulletin}\/},  {\it
  \bibinfo{volume}{112}\/}, \bibinfo{pages}{155--159}.
\bibitem[{Cruz et~al.(2015)Cruz, da~Silva \& Capretz}]{forty}
\bibinfo{author}{Cruz, S.}, \bibinfo{author}{da~Silva, F.~Q.}, \&
  \bibinfo{author}{Capretz, L.~F.} (\bibinfo{year}{2015}).
\newblock \bibinfo{title}{Forty years of research on personality in software
  engineering: A mapping study}.
\newblock {\it \bibinfo{journal}{Computers in Human Behavior}\/},  {\it
  \bibinfo{volume}{46}\/}, \bibinfo{pages}{94--113}.
\bibitem[{De~Wit(1988)}]{de1988measurement}
\bibinfo{author}{De~Wit, A.} (\bibinfo{year}{1988}).
\newblock \bibinfo{title}{Measurement of project success}.
\newblock {\it \bibinfo{journal}{International journal of project
  management}\/},  {\it \bibinfo{volume}{6}\/}, \bibinfo{pages}{164--170}.
\bibitem[{Dikert et~al.(2016)Dikert, Paasivaara \&
  Lassenius}]{dikert2016challenges}
\bibinfo{author}{Dikert, K.}, \bibinfo{author}{Paasivaara, M.}, \&
  \bibinfo{author}{Lassenius, C.} (\bibinfo{year}{2016}).
\newblock \bibinfo{title}{Challenges and success factors for large-scale agile
  transformations: A systematic literature review}.
\newblock {\it \bibinfo{journal}{Journal of Systems and Software}\/},  {\it
  \bibinfo{volume}{119}\/}, \bibinfo{pages}{87--108}.
\bibitem[{Feldt et~al.(2010)Feldt, Angelis, Torkar \& Samuelsson}]{feldt2010}
\bibinfo{author}{Feldt, R.}, \bibinfo{author}{Angelis, L.},
  \bibinfo{author}{Torkar, R.}, \& \bibinfo{author}{Samuelsson, M.}
  (\bibinfo{year}{2010}).
\newblock \bibinfo{title}{Links between the personalities, views and attitudes
  of software engineers}.
\newblock {\it \bibinfo{journal}{Information and Software Technology}\/},  {\it
  \bibinfo{volume}{52}\/}, \bibinfo{pages}{611--624}.
\bibitem[{Gren et~al.(2014)Gren, Torkar \& Feldt}]{gren1}
\bibinfo{author}{Gren, L.}, \bibinfo{author}{Torkar, R.}, \&
  \bibinfo{author}{Feldt, R.} (\bibinfo{year}{2014}).
\newblock \bibinfo{title}{Work motivational challenges regarding the interface
  between agile teams and a non-agile surrounding organization: A case study}.
\newblock In {\it \bibinfo{booktitle}{Agile Conference (AGILE), 2014}\/} (pp.
  \bibinfo{pages}{11–--15}).
\bibitem[{Gren et~al.(2015{\natexlab{a}})Gren, Torkar \& Feldt}]{gren2015seaa}
\bibinfo{author}{Gren, L.}, \bibinfo{author}{Torkar, R.}, \&
  \bibinfo{author}{Feldt, R.} (\bibinfo{year}{2015}{\natexlab{a}}).
\newblock \bibinfo{title}{Group maturity and agility, are they connected? –
  {A} survey study}.
\newblock In {\it \bibinfo{booktitle}{Proceedings of the 41st EUROMICRO
  Conference on Software Engineering and Advanced Applications (SEAA)}\/} (pp.
  \bibinfo{pages}{1–--8}).
\newblock \bibinfo{publisher}{IEEE}.
\bibitem[{Gren et~al.(2015{\natexlab{b}})Gren, Torkar \& Feldt}]{grenjss}
\bibinfo{author}{Gren, L.}, \bibinfo{author}{Torkar, R.}, \&
  \bibinfo{author}{Feldt, R.} (\bibinfo{year}{2015}{\natexlab{b}}).
\newblock \bibinfo{title}{The prospects of a quantitative measurement of
  agility: A validation study on an agile maturity model}.
\newblock {\it \bibinfo{journal}{Journal of Systems and Software}\/},  {\it
  \bibinfo{volume}{107}\/}, \bibinfo{pages}{38—--49}.
\bibitem[{Guzzo et~al.(1985)Guzzo, Jette \& Katzell}]{guzzo1985}
\bibinfo{author}{Guzzo, R.}, \bibinfo{author}{Jette, R.}, \&
  \bibinfo{author}{Katzell, R.} (\bibinfo{year}{1985}).
\newblock \bibinfo{title}{The effects of psychologically based intervention
  programs on worker productivity: A meta-analysis}.
\newblock {\it \bibinfo{journal}{Personnel Psychology}\/},  {\it
  \bibinfo{volume}{38}\/}, \bibinfo{pages}{275--291}.
\bibitem[{Hersey et~al.(2000)Hersey, Blanchard \& Johnson}]{hersey}
\bibinfo{author}{Hersey, P.}, \bibinfo{author}{Blanchard, K.}, \&
  \bibinfo{author}{Johnson, D.} (\bibinfo{year}{2000}).
\newblock {\it \bibinfo{title}{Management of organizational behavior: Leading
  human resources}\/}.
\newblock (\bibinfo{edition}{8th} ed.).
\newblock \bibinfo{address}{Upper Saddle River, N.J.}:
  \bibinfo{publisher}{Prentice Hall}.
\bibitem[{Hudson \& Fraley(2015)}]{hudson2015volitional}
\bibinfo{author}{Hudson, N.~W.}, \& \bibinfo{author}{Fraley, R.~C.}
  (\bibinfo{year}{2015}).
\newblock \bibinfo{title}{Volitional personality trait change: Can people
  choose to change their personality traits?}
\newblock {\it \bibinfo{journal}{Journal of personality and social
  psychology}\/},  {\it \bibinfo{volume}{109}\/}, \bibinfo{pages}{490}.
\bibitem[{Iivari \& Iivari(2011)}]{iivari}
\bibinfo{author}{Iivari, J.}, \& \bibinfo{author}{Iivari, N.}
  (\bibinfo{year}{2011}).
\newblock \bibinfo{title}{The relationship between organizational culture and
  the deployment of agile methods}.
\newblock {\it \bibinfo{journal}{Information and Software Technology}\/},  {\it
  \bibinfo{volume}{53}\/}, \bibinfo{pages}{509--520}.
\bibitem[{Jalali \& Wohlin(2010)}]{jala}
\bibinfo{author}{Jalali, S.}, \& \bibinfo{author}{Wohlin, C.}
  (\bibinfo{year}{2010}).
\newblock \bibinfo{title}{Agile practices in global software engineering -- {A}
  systematic map}.
\newblock In {\it \bibinfo{booktitle}{International Conference on Global
  Software Engineering (ICGSE)}\/}.
\newblock \bibinfo{organization}{IEEE}.
\bibitem[{Jalali et~al.(2014)Jalali, Wohlin \& Angelis}]{jalali2014}
\bibinfo{author}{Jalali, S.}, \bibinfo{author}{Wohlin, C.}, \&
  \bibinfo{author}{Angelis, L.} (\bibinfo{year}{2014}).
\newblock \bibinfo{title}{Investigating the applicability of agility assessment
  surveys: A case study}.
\newblock {\it \bibinfo{journal}{Journal of Systems and Software}\/},  {\it
  \bibinfo{volume}{98}\/}, \bibinfo{pages}{172--190}.
\bibitem[{Keyton(2002)}]{grupp}
\bibinfo{author}{Keyton, J.} (\bibinfo{year}{2002}).
\newblock {\it \bibinfo{title}{Communicating in groups: Building relationships
  for group effectiveness}\/}.
\newblock \bibinfo{address}{New York}: \bibinfo{publisher}{McGraw-Hill}.
\bibitem[{Korhonen(2011)}]{korhonen}
\bibinfo{author}{Korhonen, K.} (\bibinfo{year}{2011}).
\newblock \bibinfo{title}{Adopting agile practices in teams with no direct
  programming responsibility -- {A} case study}.
\newblock {\it \bibinfo{journal}{Product-Focused Software Process
  Improvement}\/},  (pp. \bibinfo{pages}{30--43}).
\bibitem[{Kotter(2007)}]{kotter2007leading}
\bibinfo{author}{Kotter, J.~P.} (\bibinfo{year}{2007}).
\newblock \bibinfo{title}{Leading change: {W}hy transformation efforts fail}.
\newblock {\it \bibinfo{journal}{Harvard Business Review}\/},  {\it
  \bibinfo{volume}{92}\/}, \bibinfo{pages}{107}.
\bibitem[{Kurapati et~al.(2012)Kurapati, Manyam \& Petersen}]{kurapati}
\bibinfo{author}{Kurapati, N.}, \bibinfo{author}{Manyam, V.}, \&
  \bibinfo{author}{Petersen, K.} (\bibinfo{year}{2012}).
\newblock \bibinfo{title}{Agile software development practice adoption survey}.
\newblock {\it \bibinfo{journal}{Agile Processes in Software Engineering and
  Extreme Programming}\/},  (pp. \bibinfo{pages}{16--30}).
\bibitem[{Lenberg et~al.(2015)Lenberg, Feldt \& Wallgren}]{lenberg2015}
\bibinfo{author}{Lenberg, P.}, \bibinfo{author}{Feldt, R.}, \&
  \bibinfo{author}{Wallgren, L.-G.} (\bibinfo{year}{2015}).
\newblock \bibinfo{title}{Behavioral software engineering: {A} definiton and
  systematic literature review}.
\newblock {\it \bibinfo{journal}{Journal of Systems and Software}\/},  {\it
  \bibinfo{volume}{107}\/}, \bibinfo{pages}{15 -- 37}.
\bibitem[{Lepp{\"a}nen(2013)}]{lepp}
\bibinfo{author}{Lepp{\"a}nen, M.} (\bibinfo{year}{2013}).
\newblock \bibinfo{title}{A comparative analysis of agile maturity models}.
\newblock In {\it \bibinfo{booktitle}{Information Systems Development}\/} (pp.
  \bibinfo{pages}{329--343}).
\newblock \bibinfo{publisher}{Springer}.
\bibitem[{McDonald \& Edwards(2007)}]{mcdonald}
\bibinfo{author}{McDonald, S.}, \& \bibinfo{author}{Edwards, H.}
  (\bibinfo{year}{2007}).
\newblock \bibinfo{title}{Who should test whom?}
\newblock {\it \bibinfo{journal}{Communications of the ACM}\/},  {\it
  \bibinfo{volume}{50}\/}, \bibinfo{pages}{66--71}.
\bibitem[{Melnik \& Maurer(2006)}]{melnik2}
\bibinfo{author}{Melnik, G.}, \& \bibinfo{author}{Maurer, F.}
  (\bibinfo{year}{2006}).
\newblock \bibinfo{title}{Comparative analysis of job satisfaction in agile and
  non-agile software development teams}.
\newblock In {\it \bibinfo{booktitle}{Extreme Programming and Agile Processes
  in Software Engineering}\/} (pp. \bibinfo{pages}{32--42}).
\newblock \bibinfo{publisher}{Springer}.
\bibitem[{Murphy \& Myors(2004)}]{murphy2004spa}
\bibinfo{author}{Murphy, K.~R.}, \& \bibinfo{author}{Myors, B.}
  (\bibinfo{year}{2004}).
\newblock {\it \bibinfo{title}{Statistical power analysis: {A} simple and
  general model for traditional and modern hypothesis tests}\/}.
\newblock (\bibinfo{edition}{2nd} ed.).
\newblock \bibinfo{address}{Mahwah, N.J.}: \bibinfo{publisher}{Lawrence
  Erlbaum}.
\bibitem[{Noll et~al.(2010)Noll, Beecham \& Richardson}]{noll2010global}
\bibinfo{author}{Noll, J.}, \bibinfo{author}{Beecham, S.}, \&
  \bibinfo{author}{Richardson, I.} (\bibinfo{year}{2010}).
\newblock \bibinfo{title}{Global software development and collaboration:
  Barriers and solutions}.
\newblock {\it \bibinfo{journal}{ACM Inroads}\/},  {\it \bibinfo{volume}{1}\/},
  \bibinfo{pages}{66--78}.
\bibitem[{Olszewska et~al.(2016)Olszewska, Heidenberg, Weijola, Mikkonen \&
  Porres}]{olszewska2016quantitatively}
\bibinfo{author}{Olszewska, M.}, \bibinfo{author}{Heidenberg, J.},
  \bibinfo{author}{Weijola, M.}, \bibinfo{author}{Mikkonen, K.}, \&
  \bibinfo{author}{Porres, I.} (\bibinfo{year}{2016}).
\newblock \bibinfo{title}{Quantitatively measuring a large-scale agile
  transformation}.
\newblock {\it \bibinfo{journal}{Journal of Systems and Software}\/},  {\it
  \bibinfo{volume}{117}\/}, \bibinfo{pages}{258--273}.
\bibitem[{Ozcan-Top \& Demirors(2013)}]{ozcan}
\bibinfo{author}{Ozcan-Top, O.}, \& \bibinfo{author}{Demirors, O.}
  (\bibinfo{year}{2013}).
\newblock \bibinfo{title}{Assessment of agile maturity models: A multiple case
  study}.
\newblock In \bibinfo{editor}{T.~Woronowicz}, \bibinfo{editor}{T.~Rout},
  \bibinfo{editor}{R.~O’Connor}, \& \bibinfo{editor}{A.~Dorling} (Eds.), {\it
  \bibinfo{booktitle}{Software Process Improvement and Capability
  Determination}\/} (pp. \bibinfo{pages}{130--141}).
\newblock \bibinfo{publisher}{Springer Berlin Heidelberg} volume
  \bibinfo{volume}{349} of {\it \bibinfo{series}{Communications in Computer and
  Information Science}\/}.
\bibitem[{Petersen \& Wohlin(2010)}]{petersen2}
\bibinfo{author}{Petersen, K.}, \& \bibinfo{author}{Wohlin, C.}
  (\bibinfo{year}{2010}).
\newblock \bibinfo{title}{The effect of moving from a plan-driven to an
  incremental software development approach with agile practices}.
\newblock {\it \bibinfo{journal}{Empirical Software Engineering}\/},  {\it
  \bibinfo{volume}{15}\/}, \bibinfo{pages}{654--693}.
\bibitem[{Ranganath(2011)}]{doingtobeing}
\bibinfo{author}{Ranganath, P.} (\bibinfo{year}{2011}).
\newblock \bibinfo{title}{Elevating teams from {`}doing{'} agile to {`}being{'}
  and {`}living{'} agile}.
\newblock In {\it \bibinfo{booktitle}{Agile Conference (AGILE), 2011}\/} (pp.
  \bibinfo{pages}{187--194}).
\bibitem[{Seger et~al.(2008)Seger, Hazzan \& Bar-Nahor}]{seger}
\bibinfo{author}{Seger, T.}, \bibinfo{author}{Hazzan, O.}, \&
  \bibinfo{author}{Bar-Nahor, R.} (\bibinfo{year}{2008}).
\newblock \bibinfo{title}{Agile orientation and psychological needs,
  self-efficacy, and perceived support: A two job-level comparison}.
\newblock In {\it \bibinfo{booktitle}{Agile Conference 2008}\/} (pp.
  \bibinfo{pages}{3--14}).
\newblock \bibinfo{organization}{IEEE}.
\bibitem[{Serrador \& Pinto(2015)}]{serrador2015does}
\bibinfo{author}{Serrador, P.}, \& \bibinfo{author}{Pinto, J.~K.}
  (\bibinfo{year}{2015}).
\newblock \bibinfo{title}{Does agile work? -- {A} quantitative analysis of
  agile project success}.
\newblock {\it \bibinfo{journal}{International Journal of Project
  Management}\/},  {\it \bibinfo{volume}{33}\/}, \bibinfo{pages}{1040--1051}.
\bibitem[{Sidky(2007)}]{sidkyphd}
\bibinfo{author}{Sidky, A.} (\bibinfo{year}{2007}).
\newblock {\it \bibinfo{title}{A structured approach to adopting agile
  practices: The agile adoption framework}\/}.
\newblock Ph.D. thesis Virginia Polytechnic Institute and State University.
\bibitem[{Simon et~al.(2000)Simon, Agazarian \& Beck}]{simon}
\bibinfo{author}{Simon, A.}, \bibinfo{author}{Agazarian, Y.}, \&
  \bibinfo{author}{Beck, A.} (\bibinfo{year}{2000}).
\newblock \bibinfo{title}{The system for analyzing verbal interaction.}
\newblock In \bibinfo{editor}{C.~Lewis} (Ed.), {\it \bibinfo{booktitle}{The
  process of group psychotherapy: Systems for analyzing change.}\/} (pp.
  \bibinfo{pages}{357--380}).
\newblock \bibinfo{address}{Washington, DC}: \bibinfo{publisher}{American
  Psychological Association}.
\bibitem[{So \& Scholl(2009)}]{so}
\bibinfo{author}{So, C.}, \& \bibinfo{author}{Scholl, W.}
  (\bibinfo{year}{2009}).
\newblock \bibinfo{title}{Perceptive agile measurement: New instruments for
  quantitative studies in the pursuit of the social-psychological effect of
  agile practices}.
\newblock In {\it \bibinfo{booktitle}{Agile Processes in Software Engineering
  and Extreme Programming}\/} (pp. \bibinfo{pages}{83--93}).
\newblock \bibinfo{publisher}{Springer}.
\bibitem[{Sundstr{\"o}m et~al.(1990)Sundstr{\"o}m, De~Meuse \&
  Futrell}]{sundstrom1990}
\bibinfo{author}{Sundstr{\"o}m, E.}, \bibinfo{author}{De~Meuse, K.}, \&
  \bibinfo{author}{Futrell, D.} (\bibinfo{year}{1990}).
\newblock \bibinfo{title}{Work teams: Applications and effectiveness.}
\newblock {\it \bibinfo{journal}{American psychologist}\/},  {\it
  \bibinfo{volume}{45}\/}, \bibinfo{pages}{120}.
\bibitem[{Teh et~al.(2012)Teh, Baniassad, Van~Rooy \& Boughton}]{teh}
\bibinfo{author}{Teh, A.}, \bibinfo{author}{Baniassad, E.},
  \bibinfo{author}{Van~Rooy, D.}, \& \bibinfo{author}{Boughton, C.}
  (\bibinfo{year}{2012}).
\newblock \bibinfo{title}{Social psychology and software teams: Establishing
  task-effective group norms}.
\newblock {\it \bibinfo{journal}{IEEE Software}\/},  {\it
  \bibinfo{volume}{29}\/}, \bibinfo{pages}{53--58}.
\bibitem[{Terracciano et~al.(2005)Terracciano, McCrae, Brant \&
  Costa~Jr}]{terracciano2005hierarchical}
\bibinfo{author}{Terracciano, A.}, \bibinfo{author}{McCrae, R.~R.},
  \bibinfo{author}{Brant, L.~J.}, \& \bibinfo{author}{Costa~Jr, P.~T.}
  (\bibinfo{year}{2005}).
\newblock \bibinfo{title}{Hierarchical linear modeling analyses of the
  {NEO-PI-R} scales in the {B}altimore longitudinal study of aging.}
\newblock {\it \bibinfo{journal}{Psychology and aging}\/},  {\it
  \bibinfo{volume}{20}\/}, \bibinfo{pages}{493}.
\bibitem[{Tolfo \& Wazlawick(2008)}]{tolfo2008}
\bibinfo{author}{Tolfo, C.}, \& \bibinfo{author}{Wazlawick, R.}
  (\bibinfo{year}{2008}).
\newblock \bibinfo{title}{The influence of organizational culture on the
  adoption of extreme programming}.
\newblock {\it \bibinfo{journal}{Journal of systems and software}\/},  {\it
  \bibinfo{volume}{81}\/}, \bibinfo{pages}{1955--1967}.
\bibitem[{Tolfo et~al.(2011)Tolfo, Wazlawick, Ferreira \& Forcellini}]{tolfo}
\bibinfo{author}{Tolfo, C.}, \bibinfo{author}{Wazlawick, R.},
  \bibinfo{author}{Ferreira, M.}, \& \bibinfo{author}{Forcellini, F.}
  (\bibinfo{year}{2011}).
\newblock \bibinfo{title}{Agile methods and organizational culture: Reflections
  about cultural levels}.
\newblock {\it \bibinfo{journal}{Journal of Software Maintenance and Evolution:
  Research and Practice}\/},  {\it \bibinfo{volume}{23}\/},
  \bibinfo{pages}{423--441}.
\bibitem[{Tuckman \& Jensen(1977)}]{tuckman}
\bibinfo{author}{Tuckman, B.}, \& \bibinfo{author}{Jensen, M.}
  (\bibinfo{year}{1977}).
\newblock \bibinfo{title}{Stages of small-group development revisited}.
\newblock {\it \bibinfo{journal}{Group \& Organization Management}\/},  {\it
  \bibinfo{volume}{2}\/}, \bibinfo{pages}{419--427}.
\bibitem[{Weinberg(1998)}]{weinberg}
\bibinfo{author}{Weinberg, G.} (\bibinfo{year}{1998}).
\newblock {\it \bibinfo{title}{The psychology of computer programming}\/}.
\newblock (\bibinfo{edition}{Silver anniversary} ed.).
\newblock \bibinfo{address}{New York}: \bibinfo{publisher}{Dorset House Pub}.
\bibitem[{Wheelan(2009)}]{wheelan2009}
\bibinfo{author}{Wheelan, S.} (\bibinfo{year}{2009}).
\newblock \bibinfo{title}{Group size, group development, and group
  productivity}.
\newblock {\it \bibinfo{journal}{Small Group Research}\/},  {\it
  \bibinfo{volume}{40}\/}, \bibinfo{pages}{247--262}.
\bibitem[{Wheelan(2013)}]{wheelan2012}
\bibinfo{author}{Wheelan, S.} (\bibinfo{year}{2013}).
\newblock {\it \bibinfo{title}{Creating effective teams: {A} guide for members
  and leaders}\/}.
\newblock (\bibinfo{edition}{4th} ed.).
\newblock \bibinfo{address}{Thousand Oaks}: \bibinfo{publisher}{SAGE}.
\bibitem[{Wheelan et~al.(2003)Wheelan, Burchill \& Tilin}]{wheelan20032}
\bibinfo{author}{Wheelan, S.}, \bibinfo{author}{Burchill, C.~N.}, \&
  \bibinfo{author}{Tilin, F.} (\bibinfo{year}{2003}).
\newblock \bibinfo{title}{The link between teamwork and patients' outcomes in
  intensive care units}.
\newblock {\it \bibinfo{journal}{American Journal of Critical Care}\/},  {\it
  \bibinfo{volume}{12}\/}, \bibinfo{pages}{527--534}.
\bibitem[{Wheelan \& Hochberger(1996)}]{wheelan}
\bibinfo{author}{Wheelan, S.}, \& \bibinfo{author}{Hochberger, J.}
  (\bibinfo{year}{1996}).
\newblock \bibinfo{title}{Validation studies of the group development
  questionnaire}.
\newblock {\it \bibinfo{journal}{Small Group Research}\/},  {\it
  \bibinfo{volume}{27}\/}, \bibinfo{pages}{143--170}.
\bibitem[{Wheelan \& Kesselring(2005)}]{wheelan2005}
\bibinfo{author}{Wheelan, S.}, \& \bibinfo{author}{Kesselring, J.}
  (\bibinfo{year}{2005}).
\newblock \bibinfo{title}{Link between faculty group development and elementary
  student performance on standardized tests}.
\newblock {\it \bibinfo{journal}{The journal of educational research}\/},  {\it
  \bibinfo{volume}{98}\/}, \bibinfo{pages}{323--330}.
\bibitem[{Wheelan \& Mckeage(1993)}]{wheelan1993}
\bibinfo{author}{Wheelan, S.}, \& \bibinfo{author}{Mckeage, R.}
  (\bibinfo{year}{1993}).
\newblock \bibinfo{title}{Developmental patterns in small and large groups}.
\newblock {\it \bibinfo{journal}{Small Group Research}\/},  {\it
  \bibinfo{volume}{24}\/}, \bibinfo{pages}{60--83}.
\bibitem[{Wheelan et~al.(1998)Wheelan, Murphy, Tsumura \& Kline}]{wheelan1998}
\bibinfo{author}{Wheelan, S.}, \bibinfo{author}{Murphy, D.},
  \bibinfo{author}{Tsumura, E.}, \& \bibinfo{author}{Kline, S.~F.}
  (\bibinfo{year}{1998}).
\newblock \bibinfo{title}{Member perceptions of internal group dynamics and
  productivity}.
\newblock {\it \bibinfo{journal}{Small Group Research}\/},  {\it
  \bibinfo{volume}{29}\/}, \bibinfo{pages}{371--393}.
\bibitem[{Wheelan \& Tilin(1999)}]{wheelan1999}
\bibinfo{author}{Wheelan, S.}, \& \bibinfo{author}{Tilin, F.}
  (\bibinfo{year}{1999}).
\newblock \bibinfo{title}{The relationship between faculty group development
  and school productivity}.
\newblock {\it \bibinfo{journal}{Small group research}\/},  {\it
  \bibinfo{volume}{30}\/}, \bibinfo{pages}{59--81}.
\bibitem[{Wheelan(2005)}]{wheelandev}
\bibinfo{author}{Wheelan, S.~A.} (\bibinfo{year}{2005}).
\newblock {\it \bibinfo{title}{Group processes: {A} developmental
  perspective}\/}.
\newblock (\bibinfo{edition}{2nd} ed.).
\newblock \bibinfo{address}{Boston}: \bibinfo{publisher}{Allyn and Bacon}.
\bibitem[{Whitworth \& Biddle(2007)}]{whit}
\bibinfo{author}{Whitworth, E.}, \& \bibinfo{author}{Biddle, R.}
  (\bibinfo{year}{2007}).
\newblock \bibinfo{title}{The social nature of agile teams}.
\newblock In {\it \bibinfo{booktitle}{Agile Conference (AGILE), 2007}\/} (pp.
  \bibinfo{pages}{26--36}).
\newblock \bibinfo{organization}{IEEE}.
\bibitem[{Williams(2012)}]{williams}
\bibinfo{author}{Williams, L.} (\bibinfo{year}{2012}).
\newblock \bibinfo{title}{What agile teams think of agile principles}.
\newblock {\it \bibinfo{journal}{Communications of the ACM}\/},  {\it
  \bibinfo{volume}{55}\/}, \bibinfo{pages}{71--76}.
\bibitem[{Yu(2014)}]{seovercoming}
\bibinfo{author}{Yu, L.} (\bibinfo{year}{2014}).
\newblock {\it \bibinfo{title}{Overcoming Challenges in Software Engineering
  Education: Delivering Non-Technical Knowledge and Skills}\/}.
\newblock Advances in Higher Education and Professional Development.
\newblock \bibinfo{address}{Hershey, Pennsylvania}: \bibinfo{publisher}{IGI
  Global}.

\end{thebibliography}
\newpage

\end{document}